\documentclass[aps,preprint,a4paper,amsmath,amssymb,floatfix]{revtex4}
\usepackage{color,graphicx}
\usepackage{dcolumn}
\usepackage{bm}
\newcommand{\Op}[1]{{\boldsymbol{\mathrm{\hat{#1}}}}}

\newcommand{\beq}{\begin{equation}}
\newcommand{\eeq}{\end{equation}}
\newcommand{\beqar}{\begin{eqnarray}}
\newcommand{\eeqar}{\end{eqnarray}}
\newcommand{\bea}{\begin{eqnarray}}
\newcommand{\eea}{\end{eqnarray}}
\newcommand{\bcen}{\begin{center}}
\newcommand{\ecen}{\end{center}}

\begin{document}

\title{The Quantum Harmonic Otto Cycle}
\author{ Ronnie Kosloff $^{1,*,\dagger}$ and Yair Rezek $^{2,3\dagger}$}

\affiliation{%
$^{1}$ \quad Institute of Chemistry, The Hebrew University, Jerusalem 91904, Israel\\
$^{2}$ \quad Computer Science Department, Technion, Haifa 32000, Israel; yair.rezek@mail.huji.ac.il\\
$^{3}$ \quad Tel Hai College, Kiryat Shmona, 1220800 Israel}

\begin{abstract}
The quantum Otto cycle serves as a bridge between the macroscopic world of heat engines
and the quantum regime of thermal devices composed from a single element. 
We compile recent studies of the quantum Otto cycle with a harmonic oscillator as a working medium. 
This~model has the advantage that it  is analytically trackable. In addition, an experimental realization has been achieved, employing
a single ion in a harmonic trap. The review is embedded in the field of quantum thermodynamics and quantum open systems.
The basic principles of the theory are explained by a~specific example illuminating the basic definitions of work and heat. 
The~relation between quantum observables and the state of the system is emphasized. 
The dynamical description of the cycle is based on a completely positive map  formulated as a propagator for each stroke of the engine.
Explicit solutions for these propagators are described on a vector space of quantum thermodynamical observables. These solutions
which employ different assumptions and techniques are compared. 
The  tradeoff between power and efficiency is the focal point of finite-time-thermodynamics. 
The~dynamical model enables the study of finite time cycles limiting time on the {\em adiabatic}  and the thermalization times. 
Explicit  finite time solutions are found which are frictionless (meaning that no coherence is generated), and are also known as shortcuts to adiabaticity.
The~transition from frictionless to sudden {\em adiabats} is characterized by a non-hermitian degeneracy in the propagator.
In addition, the influence  of noise on the control is illustrated.
These results are used to close the cycles either as engines or as refrigerators. The properties of the limit cycle are described.
Methods to optimize the power by controlling the thermalization time are also introduced. 
At high temperatures, the Novikov--Curzon--Ahlborn 
efficiency at maximum power is obtained.
The~sudden limit of the engine which allows finite power at zero cycle time is shown.
The refrigerator cycle is described within the frictionless limit, with  emphasis on the cooling rate when the cold bath temperature
approaches zero.
\end{abstract}

\maketitle
\tableofcontents
\section{Introduction} 
\label{sec:introduction}

Quantum thermodynamics is devoted to the link between thermodynamical processes and their quantum origin. 
Typically, thermodynamics is applied to large macroscopic entities. Therefore,  to what extent is it possible to miniaturize. 
Can thermodynamics be applicable to the level of  a single quantum device? 
We will address this issue in the tradition of thermodynamics, by learning from an~example: Analysis of the performance of a heat engine \cite{carnot1872reflexions}.
To this end, we review recent progress in the study of the quantum harmonic oscillator as a working medium of a thermal device. 
The engine composed of a single harmonic oscillator connected to a hot and cold bath is an ideal analytically solvable model for a quantum thermal device. 
It has therefore been studied extensively and inspired experimental realisation.
Recently, a single ion heat engine with an effective harmonic trap frequency has been experimentally realised  \cite{Rossnagel325}. 
This device could roughly be classified as a reciprocating Otto engine operating
by periodically modulating the trap frequency.

Real heat engines operate far from reversible conditions.
Their performance  resides between the point of maximum efficiency and maximum power.
This has been the subject of finite time thermodynamics \cite{andresen1977,salamon01}. The topic has been devoted
to the irreversible cost of operating at finite power.
Quantum engines add a twist to the subject, as they naturally incorporate dynamics into thermodynamics \cite{alicki1979quantum,k24}.

Quantum heat engines can be classified either as continuous or reciprocating. 
The prototype of a~continuous engine is the three-level amplifier pioneered by Scovil and Schulz-DuBois \cite{scovil1959three}.
This~device is simultaneously  coupled to three input currents. It is therefore termed a tricycle, and can operate either as an engine or as a refrigerator. 
A review of continuous quantum heat engines has been published recently \cite{k289} and therefore is beyond the scope of this review. 

Reciprocating engines are classified according to their sequence of strokes. 
The most studied cycles are Carnot  \cite{k85,k87,bender,lloyd,esposito2010efficiency} and Otto cycles \cite{k116,quan07,heJ09,henrich2007,agarwal2013quantum,zhang2014prl}.
The quantum Otto cycle  is easier to analyze, and therefore it became the primary example of a reciprocating quantum heat engine.  
The pioneering studies of quantum reciprocating engines employed a two-level system---a qubit---as a working \mbox{medium \cite{k85,k87,k116,he02}}. 
The performance  analysis of the quantum versions of reciprocating engines exhibited an amazing resemblance to macroscopic counterparts.
For example, the efficiency at maximum power of the quantum version of the endoreversiable engine converges at high
temperature to the Novikov--Curzon--Ahlborn macroscopic Newtonian model predictions \cite{novikov1958efficiency,curzon75}.
The deviations were even small at low temperature, despite the fact that the heat transport law was different \cite{k85}.
The only quantum feature that could be identified was related to the discrete structure of the energy levels.

Heat engines with quantum features require a more complex working medium than a single qubit weakly coupled to a heat bath. 
This complexity is required to obtain quantum analogues of friction and heat leaks.
A prerequisite for such phenomena is that the external control part of the Hamiltonian does not commute with the internal part.
This generates  quantum non-adiabatic phenomena which lead to friction \cite{k176}.
A working medium composed of a quantum harmonic oscillator has sufficient complexity to represent generic phenomena,
but can still be amenable to analytic analysis
~\cite{k221}. 

The quantum Otto cycle is a primary example of the emerging field of quantum thermodynamics. The quest is to
establish the similarities and differences in applying thermodynamic reasoning up to the level of a single quantum entity.
The present analysis is based on the theory of quantum open systems~\cite{k281,gemmer2009}.  
A dynamical description based on the weak system bath coupling has been able to establish 
consistency between quantum mechanics and the laws of thermodynamics \cite{k281}. 
These~links allow work and heat to be defined  in the quantum regime \cite{talkner2016}. This framework is sufficient for the present~analysis.

In the strong coupling regime where the partition between system and bath is not clear, the connection to thermodynamics
is not yet established---although different approaches have been suggested \cite{seifert2016first,carrega2016energy}.
A different approach to quantum thermodynamics termed quantum thermodynamics resource theory follows ideas from quantum information resource theory, establishing a set of rules 
~\cite{goold16,vinjanampathy2015}. We will try to show 
how this approach can be linked to the Otto cycle under~analysis.

\section{The Quantum Otto Cycle}
\label{sec:otto}

Nicolaus August Otto invented a reciprocating four stroke engine in 1861,
and won a gold medal in the 1867 Paris world fair \cite{otto}.
The basic components of the engine are hot and cold reservoirs, a~working medium, 
and a mechanical output device.
The cycle of the engine is defined by four~segments:
\begin{enumerate}
\item{The hot {\em isochore}: heat is transferred from the hot bath to the working medium without volume~change.}
\item{The power {\em adiabat}: the working medium expands, producing work, while isolated from the hot and cold reservoirs.}
\item{The cold {\em isochore}: heat is transferred from the working medium to the cold bath without \mbox{volume change}.}
\item{The compression {\em adiabat}: the working medium is compressed, consuming power while isolated from the 
hot and cold reservoirs, closing the cycle.}
\end{enumerate}

Otto  determined that the efficiency $\eta$ of the cycle is limited to 
$\eta_o \le 1 -(\frac{V_h}{V_c})^{\frac{C_p}{C_v}-1}$, 
where $V_{c/h}$ and $T_{c/h}$ are the volume and temperature of the working medium 
at the end of the hot and cold {\em isochores}, respectively. 
$C_p$ and $C_v$ are the heat capacities under constant pressure and constant volume \cite{callen}. 
As~expected, Otto efficiency is always smaller than the efficiency
of the Carnot cycle $\eta_o \le \eta_c= 1  - \frac{T_c}{T_h}$.

The first step in learning from  an example is to establish a quantum version of the Otto cycle.
This is carried out by seeking  analogues for each segment of the cycle. 
What makes the approach unique is that it is applicable to a small quantum system such as a single atom
in a harmonic trap.
The~description is embedded in the theory of open quantum systems.
Each of these segments is defined by a completely positive (CP) propagator \cite{kraus71} describing the change
of state  in the working medium: $\hat \rho_f  = {\cal U}_{i \rightarrow f} \hat \rho_i$, where the density operator $\hat \rho$ describes the state
of the working medium.

The quantum engine Otto cycle is therefore described as:
\begin{enumerate}
\item{The hot {\em isochore}: heat is transferred from the hot bath to the working medium without  change in the external parameter $\omega_h$.
The stroke is described by the propagator ${\cal U}_h$.}
\item{The expansion {\em adiabat}: the working medium reduces its energy scale. The harmonic frequency changes from $\omega_h $ to $\omega_c$, 
with $\omega_h > \omega_c$, producing work while isolated from the hot and cold reservoirs. The stroke is described by the propagator ${\cal U}_{hc}$.}
\item{The cold {\em isochore}: heat is transferred from the working medium to the cold bath without change in the external parameter $\omega_c$.
The stroke is described by the propagator ${\cal U}_c$.}
\item{The compression {\em adiabat}: the working medium increases its energy scale. 
The harmonic frequencies increase from $\omega_c $ to $\omega_h$,
consuming power while isolated from the hot and cold reservoirs. The stroke is described by the propagator ${\cal U}_{ch}$.}
\end{enumerate}

The cycle propagator becomes the product of the segment propagators:
\begin{equation}
 {\cal U}_{cyc}={\cal U}_{ch}{\cal U}_c {\cal U}_{hc}{\cal U}_h\;.
 \label{eq:cycprop}
 \end{equation}
 
\begin{figure}
\center{\includegraphics[height=9cm]{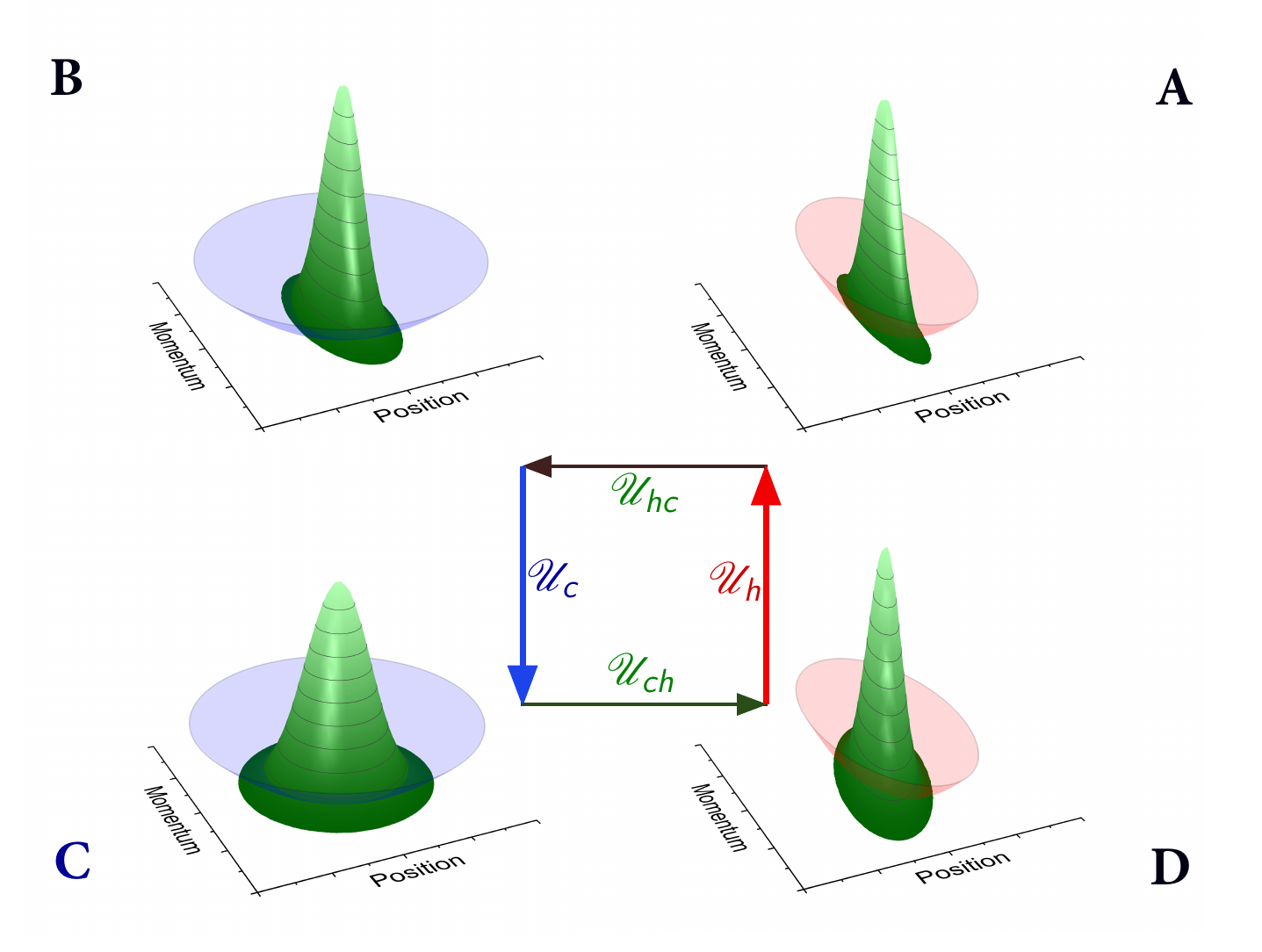}} 
\caption{Otto cycle in phase space.  The blue and red bowls represent the energy value in 
position and momentum. The compression ratio is $ {\cal C}=\frac{\omega_h}{\omega_c} =2 $. 
Expansion {\em adiabat} $A \rightarrow B$. Cold {\em isochore} $B \rightarrow C$.
Compression {\em adiabat} $C \rightarrow D$. Hot {\em isochore} $ D \rightarrow A$.
The Wigner distribution in phase space is shown in green. The state in  $A$ is a thermal
equilibrium state with the hot bath temperature. The state in $B$ is squeezed with respect to the cold bath frequency $\omega_c$. 
The state in $C$ is an equilibrium state with the cold bath temperature. The state in $D$ shows position momentum correlation $\langle \Op C \rangle \ne 0$.
} 
\label{fig:otto4}   
\end{figure}

The cycle propagator is a completely positive (CP) map of the state of the working medium \cite{kraus71}.
The order of  propagators is essential, since the segment propagators do not commute; for example, $[{\cal U}_h,{\cal U}_{hc}] \ne 0$.
The non-commuting property of the segment propagators is not an exclusive quantum property. 
It is also present  in stochastic descriptions of the engine
where the propagators operate on a~vector of populations of the energy eigenvalues.
Nevertheless, it can have a quantum origin for engines with  propagators with small action
\cite{k299,k306}.
The same operators but with different parameters (such as different frequencies) can be used to describe an Otto refrigeration cycle.
Figure \ref{fig:otto4} shows a~schematic representation of the Otto cycle in phase space.

In the adiabatic limit when the population $N$ stays constant in the expansion and compression segments, the work per cycle
becomes:
\begin{equation}
{\cal W}_{cyc} = \hbar \Delta \omega \Delta N\;,
\label{eq:work}
\end{equation}
where $\Delta \omega = \omega_h-\omega_c$ and $\Delta N$ is the population difference. $\Delta N=\Delta N_c =\Delta N_h$, since the cycle is periodic.
Under these conditions, the efficiency becomes:
\begin{equation}
\eta_o = 1 -\frac{\omega_c}{\omega_h} \le \eta_c\;,
\label{eq:adef}
\end{equation}
where $\eta_c$ is the Carnot efficiency $\eta_c=1-\frac{T_c}{T_h}$. At this stage, it is also useful to define the compression ratio
${\cal C}=\frac{\omega_h}{\omega_c}$.

\subsection{Quantum Dynamics of the Working Medium}

The quantum analogue of the Otto cycle requires a dynamical description of 
the working medium, the power output, and the heat transport mechanism.

A particle in a harmonic potential will constitute our working medium.
This choice is amenable to analytic solutions and has sufficient complexity to serve as a generic example. 
Even a single specimen is sufficient to  realize the operation of an engine.

We can imagine a single particle in a harmonic trap $V(Q)=\frac{k}{2}Q^2$.
Expansion and compression of the working medium is carried out by externally controlling the trap parameter
$k(t)$. The energy of the particle is represented by the Hamiltonian operator:
\begin{equation}
\Op H ~~=~~ \frac{1}{2m}\Op P^2 ~+ ~ \frac{k(t)}{2} {\Op Q^2}~,
\label{eq:hamils}
\end{equation}
where $m$ is the mass of the system and $\Op P$ and $\Op Q$ are the momentum and position
operators. All~thermodynamical quantities will be intensive; i.e., normalized to the number of particles.
In the macroscopic Otto engine, the internal energy of the working medium
during the adiabatic expansion is inversely proportional to the volume.
In the harmonic oscillator, the energy is linear in
the frequency $\omega(t) = \sqrt{k(t)/m}$ \cite{boyer03}. This therefore plays the role of inverse volume $\frac{1}{V}$. 

The Hamiltonian (\ref{eq:hamils}) is the generator of the evolution on the adiabatic segments.
The frequency $\omega$ changes from $\omega_h$ to $\omega_c$ in a time period $\tau_{hc}$ in the power {\em adiabat} ($\omega_h > \omega_c$)
and from $\omega_c$ to $\omega_h$ in a~period $\tau_{ch}$ in the compression {\em adiabat}. 
The dynamics of the state $\hat \rho$ during the adiabatic segments  
is unitary and is the solution of the Liouville von Neumann equation \cite{vNeumann}:
\begin{equation}
 \frac{d}{dt}{\hat \rho}(t) ~~=~~ -\frac{i}{\hbar} [\Op H(t),\hat \rho(t)]~,
 \label{eq:lvn} 
\end{equation}
where $\Op H$ is time dependent during the evolution. Notice that $[ \Op H (t),\Op H(t')] \neq 0$, since
the kinetic energy does not commute with the varying potential energy. This is the origin of quantum friction \cite{k176,plastina2014}.
The formal solution to Equation (\ref{eq:lvn}) defines the propagator:
\begin{equation}
\hat \rho(t) ~=~ {\cal U}(t) \hat \rho (0) ~=~ \Op U \hat \rho (0) \Op U^{\dagger}~, 
\label{eq:prop1}
\end{equation}
where $\Op U $ satisfies the equation:
\begin{equation}
i \hbar \frac{d}{dt} \Op U ~=~ \Op H (t) \Op U
\label{eq:u}
\end{equation}
with the initial condition $\Op U(0) = \Op I$.

The dynamics on the hot and cold {\em isochores} is a thermalization process 
of the working medium with a bath at temperature $T_h$ or $T_c$. The  dynamics
is of an open quantum system, where the working medium is described explicitly
and the bath implicitly \cite{lindblad76,gorini1976completely,breuer02}: 
\begin{equation}
 \frac{d}{dt} {\hat \rho}(t) ~~=~~ -\frac{i}{\hbar} [\Op H,\hat \rho]+{\cal L}_D (\hat \rho)~,
 \label{eq:lvn2} 
\end{equation}
where ${\cal L}_D$ is the dissipative
term responsible for driving the working medium to thermal equilibrium,
while the Hamiltonian $\Op H = \Op H(\omega_{h/c})$ is static.
The equilibration is not complete in typical operating conditions,  since only a finite time $\tau_h$ or $\tau_c$ is allocated
to the hot or cold {\em isochores}. The dissipative ``superoperator'' ${\cal L}_D$ must conform to Lindblad's form for a Markovian evolution \cite{lindblad76,gorini1976completely}, and for the harmonic oscillator can be expressed as \cite{ingarden75,louisell,lindblad1976brownian}:
\begin{equation}
\mathcal{L}_{D}(\hat \rho)
~~=~~k_{\uparrow}(\Op a^{\dagger}\hat \rho \Op {a}~-~\frac{1}{2}\{\Op {a}\Op {a}^{\dagger},\hat \rho\})+
k_{\downarrow}(\Op{a}\hat \rho \Op {a}^{\dagger}-\frac{1}{2}\{\Op{a}^{\dagger}\Op{a},\hat \rho\})\;,
\label{eq:dissipative}
\end{equation}
where anticommutator $\{ \Op A ,\Op B\} \equiv \Op A \Op B+\Op B \Op A$. $k_{\uparrow}$ and $k_{\downarrow}$ 
are heat conductance rates obeying detailed balance  
$\frac{k_{\uparrow}}{k_{\downarrow}} = e^{-\frac{\hbar \omega}{k_b T}}$, and $T$ is either $T_h$ or
$T_c$. The operators $\Op a^{\dagger}$ and $\Op a$ are the raising and lowering operators, respectively. 
Notice that they are different in the hot and cold {\em isochores}, since
$\Op a = \frac{1}{\sqrt 2} (\sqrt{\frac {m \omega}{\hbar}} \Op Q +i \sqrt{\frac{1 }{ \hbar m \omega}}\Op P)$
depends on $\omega$.  Formally for the {\em isochore} ${\cal U}_{h/c}=\exp( {\cal L} t)$ where ${\cal L} = -i /\hbar [\Op H, \cdot ]+{\cal L}_D$.

Equation (\ref{eq:dissipative}) is known as a quantum Master equation \cite{breuer02} or  L-GKS \cite{lindblad76,gorini1976completely}. 
It is an example of a reduced description
where the dynamics of the working medium is sought explicitly while the baths
are described implicitly by two parameters: the heat conductivity 
$\Gamma=k_{\downarrow}-k_{\uparrow}$ and
the bath temperature $T$. The Lindblad form of Equation (\ref{eq:dissipative}) guarantees that the density operator of the extended system (system + bath) remains positive (i.e., physical) \cite{lindblad76}. Specifically, 
for the harmonic oscillator, Equation (\ref{eq:dissipative}) has been derived from first principles by many authors \cite{braun1977quantum,um2002quantum,louisell,gardiner2004quantum,carmichael2009open}.

To summarize, the quantum model of the Otto cycle is composed of a working fluid of harmonic oscillators (\ref{eq:hamils}). 
The power stroke is modeled by the Liouville von Neumann equation (\ref{eq:lvn}), while the heat transport via a Master equation (\ref{eq:lvn2}) and (\ref{eq:dissipative}). 

\section{Quantum Thermodynamics}
\label{sec:thermo}

Thermodynamics is notorious for its ability to describe a process employing an extremely 
small number of variables. In scenarios where systems are far from thermal equilibrium, further variables have to be added.
The analogue description in quantum thermodynamics is based on a minimal set of quantum expectations 
$\langle \Op X_n \rangle$,  where 
$\langle \Op X_n \rangle = Tr \{\Op X_n \hat \rho \}$. 
 The dynamics of this set is generated by the Heisenberg equations of motion
\begin{equation}
 \frac{d}{dt}{\Op X}  ~~=~~ \frac{\partial \Op X}{\partial t}~+~\frac{i}{\hbar} [\Op H,\Op X]+{\cal L}^*_D (\Op X)~,
 \label{eq:heisenberg} 
\end{equation}
where the first term addresses an explicitly time-dependent set of operators, $\Op X(t)$.

The dynamical approach to quantum thermodynamics \cite{k281}
seeks the relation between thermodynamical laws and their quantum origin.

{\it {The first law of thermodynamics}} 
is equivalent to the energy balance relation.
The energy expectation $E$ is obtained when $\Op X = \Op H$; i.e.,  $E = \langle \Op H \rangle $.
The quantum energy partition defining the first law of thermodynamics,
$d E = d {\cal W}+ d {\cal Q}$, is obtained by inserting  $\Op H $ into (\ref{eq:heisenberg})  \cite{sphon,alicki1979quantum,k24,k281}:
\begin{equation}
 \frac{d}{dt} { E} ~~=~~ \dot{ \cal W} ~+~ \dot {\cal Q }
 ~~=~~\langle~\frac{\partial \Op H}{\partial t}~\rangle~~+~~ \langle~{\cal L}^*_D (\Op H) ~\rangle.
 \label{eq:firstlaw} 
\end{equation}

The power is identified as $${\cal P} ~=~ \dot{ \cal W}~=~ \langle~\frac{\partial \Op H}{\partial t}~\rangle.$$ 

The heat exchange rate becomes 
$$\frac{d}{dt}{\cal Q}=\langle~{\cal L}^*_D (\Op H) ~\rangle .$$ 

The analysis of the Otto cycle benefits from the simplification that power is produced or consumed
only on the {\em adiabats} and heat transfer takes place only on the {\em isochores}.

The thermodynamic state of a system is fully determined by the thermodynamical 
variables. Statistical thermodynamics adds the prescription that the state 
is determined by the maximum entropy condition
subject to the constraints set by the thermodynamical observables 
\cite{janes57a,janes57b,katz67}. Maximizing the von Neumann entropy \cite{vNeumann}
\begin{equation}
S_{VN}=-k_B Tr\{\hat {\rho}\ln(\hat {\rho})\}
\label{eq:vnentropy}
\end{equation}
subject to the energy constraint leads to thermal equilibrium \cite{katz67}
\begin{equation}
\hat \rho_{eq}~~=~~ \frac{1}{Z} e^{-\frac{\Op H}{k_B T} }~,
\label{eq:req}
\end{equation}
where $k_B$ is the Boltzmann constant and $Z=Tr\{e^{-\frac{\Op H}{k_B T}} \}$ is the partition function. 

In general, the state of the working medium  is not in thermal equilibrium. 
In order to generalize the canonical form (\ref{eq:req}), additional observables 
are required to define the state of the system.
The~maximum entropy state subject to this set of  observables   \cite{alhassid,jaynes1957a}
$\langle \Op X_j \rangle = tr \{ {\Op X_j} \hat \rho \}$ becomes
\begin{equation}
\hat \rho ~~=~~  \frac{1}{Z}\exp\left( \sum_j ~\beta_j \Op X_j  \right) ,
\label{eq:state}
\end{equation}
where $\beta_j$ are Lagrange multipliers. 
The generalized canonical form of (\ref{eq:state}) is meaningful only if 
the state can be cast in the canonical form during the entire cycle of the engine,
leading to $\beta_j =\beta_j(t)$. This~requirement is called { canonical invariance} \cite{wise}.
It implies that if an initial state belongs to the canonical class of states, it will remain in this class throughout the cycle.

A necessary condition for canonical invariance is that the set
of operators $\Op X$ in (\ref{eq:state}) is closed under the dynamics generated by the equation of motion.
If this condition is satisfied, then the state of the system can be reconstructed
from a small number of quantum observables $\langle \Op X_j \rangle(t)$. These become  the thermodynamical observables, 
since they  define the state under the maximum entropy principle.

The condition for canonical invariance on the unitary part of the evolution taking place
on the {\em adiabats} is as follows: if the Hamiltonian is a linear combination 
of the operators in the set $\Op H(t)= \sum_m h_m \Op X_m~$ ($h_m(t)$ are expansion coefficients),
and the set forms a closed Lie algebra  $[\Op X_j,\Op X_k]=\sum_l C^{jk}_l \Op X_l$
(where $C^{jk}_l$ is the structure factor of the Lie algebra), then the set $\Op X $
is closed under the evolution \cite{weinorman63}. 

For a closed Lie algebra, the generalized Gibbs state Equation (\ref{eq:state}) can always be written in a~product~form:
\begin{equation}
\hat \rho ~~=~~  \prod_k e^{ \lambda_k \Op X_k }~,
\label{eq:prod}
\end{equation}
where there is a one-to-one relation between $\lambda$ and $\beta$, depending on the order of the product form.
Multiplying the equation of motion by $\hat \rho^{-1}$ leads to $\frac{d}{dt} \hat \rho \hat \rho^{-1} = {\cal L}( \hat \rho) \hat \rho^{-1}$.
Using the product form and the Backer--Housdorff relation, the l.h.s. $\frac{d}{dt} \hat \rho \hat \rho^{-1}$ 
decomposes to a linear combination of the operator algebra. 
This is also true for the r.h.s $[ \Op H, \hat \rho] \hat \rho^{-1}$, which also becomes a linear combination of the operator algebra.
Comparing both sides of the equation of motion, one obtains a set of coupled differential
equations for the coefficients $\lambda_k$. Their solution guarantees  that canonical invariance prevails \cite{alhassid}. 

For the harmonic Otto cycle, the set of the operators $\Op P^2$, $\Op Q^2$, and $\Op D =  \frac{1}{2}(\Op Q \Op P + \Op P \Op Q)$
form a closed Lie algebra. Since the Hamiltonian is a linear combination of the first two operators
of the set ($\Op P^2$ and $\Op Q^2$), canonical invariance will prevail on the adiabatic segments.

On the {\em isochores}, the set of operators also has to be closed to the operation of  ${\cal L}_D$. 
The set $\Op P^2$, $\Op Q^2$,~and $\Op D$ is  closed to ${\cal L}_D$, defined by (\ref{eq:dissipative}).
For canonical invariance of $\hat \rho$, ${\cal L}_D \hat \rho \hat \rho^{-1} $ should also be a~linear combination of
operators in the algebra.
For the harmonic working medium and ${\cal L}_D$ defined in (\ref{eq:dissipative}), this condition is fulfilled.
As a result, canonical invariance with the set of operators $\Op P^2$, $\Op Q^2$, and $\Op D =  \frac{1}{2}(\Op Q \Op P + \Op P \Op Q)$
prevails for the whole cycle \cite{k221}.

The significance of canonical invariance is that  a solution of the operator dynamics allows the reconstruction of the state
of the working medium during the whole cycle. As a result, all dynamical quantities become functions 
of a very limited set of thermodynamic quantum observables $\langle \Op X_j \rangle$.
The~choice of  a set of operators $\{ \Op X_j \}$ should reflect the most essential thermodynamical
variables. The~operator algebra forms a vector space with the scalar product $ \left( \Op X_j \cdot \Op X_k \right) = Tr \{ \Op X_j^{\dagger} \Op X_k \} $.
This vector space will be used to describe the state $\hat \rho$ and define the cycle propagators ${\cal U}_l$. 
This description is a~significant
reduction in the dimension of the propagator ${\cal U}$ from $N^2$, where $N$ is the size of Hilbert space to $M$ the size of the 
operator algebra.

Explicitly,  variables with  thermodynamical significance are chosen for the harmonic oscillator. 
These variables are time-dependent and describe the current state of the working medium:
\begin{itemize}
\item{The Hamiltonian 
$\Op H(t)~~=~\frac{1}{2 m} \Op P^2+\frac{1}{2}m \omega(t)^2 \Op Q^2$.} 
\item{The Lagrangian
$ \Op L(t) = \frac{1}{2 m} \Op P^2-\frac{1}{2}m \omega(t)^2 \Op Q^2$.}
\item{The position momentum correlation
$~\Op C(t)= \frac{1}{2} \omega(t) (\Op Q \Op P + \Op P \Op Q) =\omega(t) \Op D$.}
\end{itemize} 

\begin{figure}
\center{\includegraphics[height=8.5cm]{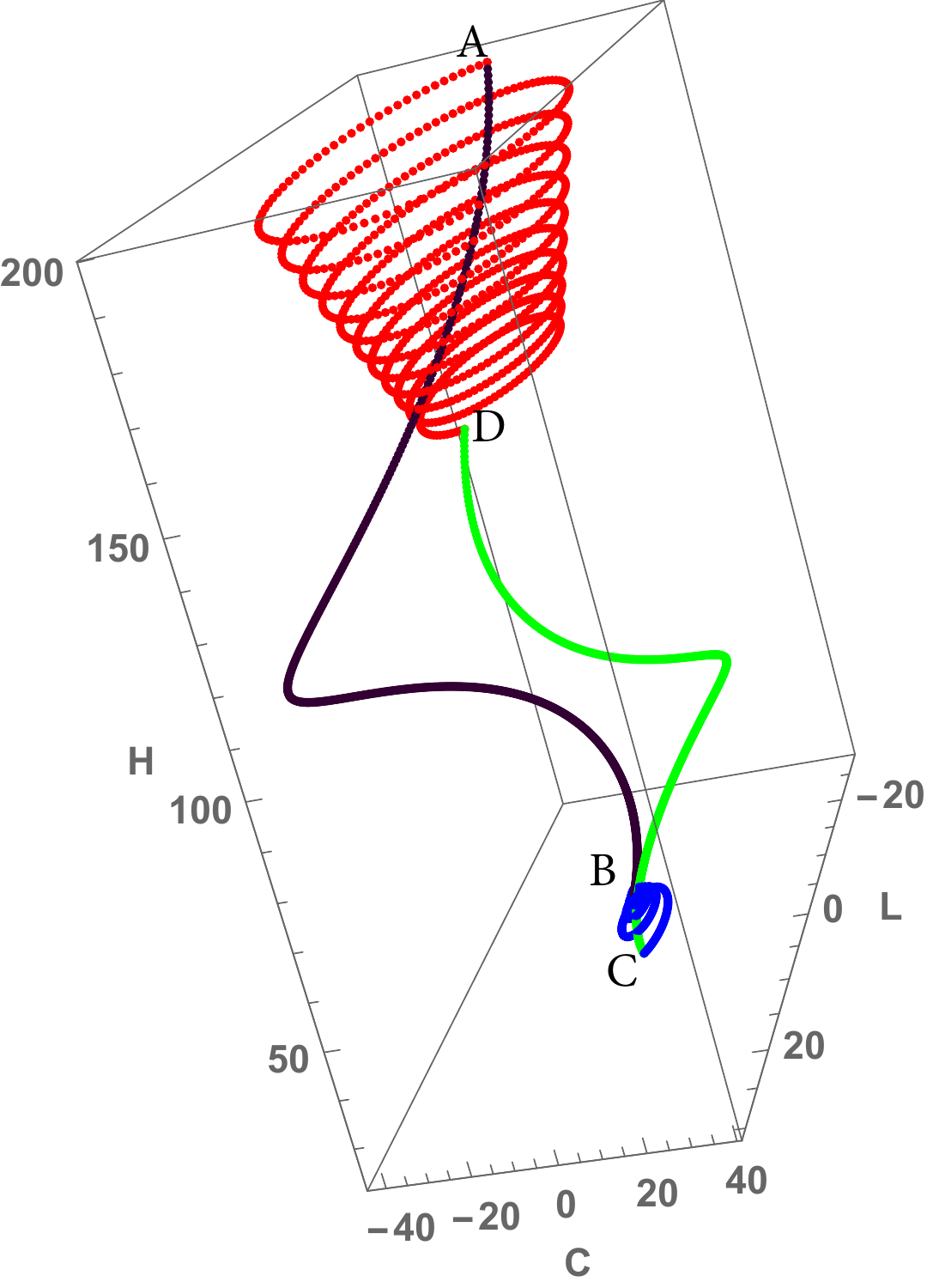}} 
\caption{Otto refrigeration cycle displayed in the  thermodynamical variables $\Op H, \Op L, \Op C$. 
When the working medium is in contact with a hot bath, the system exhausts heat  and equilibrates, 
spiralling downward from a high $\langle \Op H \rangle$ (energy) value and towards zero correlation $\langle \Op C \rangle $ and Lagrangian $\langle \Op L \rangle$ (i.e., towards thermal equilibrium). The hot {\em ishochore} is marked by the red dotted line $A \rightarrow  D$. 
On the expansion {\em adiabat},  the system spirals downwards, losing energy as it cools down---marked by the green line $D \rightarrow C$. 
It then spirals upwards (blue line), gaining energy from the cold bath $C \rightarrow B$. In addition , it spirals
~towards zero $\langle \Op C \rangle$ 
and $\langle \Op L \rangle$. Then, the compression {\em adiabat} (black line) takes it back to the top of the hot  (red) spiral $B \rightarrow A$.
} 
\label{fig:HLC}   
\end{figure}

These operators are linear combinations of the same Lie algebra as $\Op Q^2, \Op P^2 $, and $\Op D$. 
A typical cycle in terms of these variables is shown in Figure \ref{fig:HLC}.

In the algebra of operators, a special place can be attributed to 
the Casimir operator $\Op G$. This~Casimir commutes with all the operators in the algebra \cite{casimir1931,perelomov1968casimir}.
Explicitly, it becomes:
\begin{equation}
\Op G = \frac{ \Op H^2 - \Op L^2-\Op C^2}{ \hbar^2 \omega^2}
=
\frac{-\left(\Op{P}\Op{Q}+\Op{Q}\Op{P}\right)^2+2 \Op{P}\Op P \Op{Q}\Op Q+2 \Op{Q}\Op Q\Op{P}\Op P}{4 \hbar ^2}~.
\label{eq:casimir}
\end{equation}

Since $[\Op H, \Op G]=0$, $\Op G$ is constant under the evolution of the unitary segments generated by 
$\Op H$. The~Casimir for the harmonic oscillator
is  a positive operator with a minimum value determined by the uncertainty relation: 
$\langle \Op G \rangle \ge  \frac{1}{4}$ \cite{boldt2013casimir}. 

A related invariant to the dynamics is the Casimir companion \cite{boldt2013casimir}, which for the harmonic oscillator is defined as:
\begin{equation}
{\cal X} = \frac{ \langle \Op H \rangle ^2 - \langle \Op L \rangle^2-\langle \Op C \rangle^2}{ \hbar^2 \omega^2}~.
\label{eq:casimircomp}
\end{equation}
Combining Equations (\ref{eq:casimir}) and (\ref{eq:casimircomp}), an additional invariant to the dynamics can be defined:
\begin{equation}
\frac{1}{\hbar^2 \omega^2}\left(Var (\Op H) - Var (\Op L) -Var (\Op  C ) \right)=  const~,
\label{eq:var}
\end{equation}
where $Var (\Op A) = \langle \Op A^2 \rangle -\langle \Op A \rangle^2 $.

Coherence is an important quantum feature. 
The coherence is characterised by  the deviation of the state of the system
from being diagonal in energy \cite{k108,girolami2014observable}, and it can be defined as:
\begin{equation} 
{\cal C}o= \frac{1}{\hbar \omega}\sqrt{ \langle \Op L^2\rangle + \langle \Op C^2\rangle }~.
\label{eq:coherence}
\end{equation} 

From Equation (\ref{eq:casimir}), we can deduce that increasing coherence has a cost in energy $\Delta E = \hbar \omega {\cal C}o$.
 
For the closed algebra of operators, the canonical state of the system $\hat \rho$ can be cast into the
product form \cite{k221,naudts2011bch}. 
This state $\hat \rho$ is defined by the parameters $\beta$, $\gamma$, and $\gamma^*$:
\begin{equation}
\hat {\rho}~~=~~
\frac{1}{Z}e^{\gamma {\Op {a}}^{2}}e^{-\beta {\Op H}}e^{\gamma^{*}{\Op {a^{\dagger}}}^{2}}~,
\label{eq:product}
\end{equation}
where $ \Op {H}=\frac{\hbar\omega}{2}({\Op a}{\Op a^{\dagger}}+{\Op a^{\dagger}}{\Op a}) $,
$\Op {C}  =  -i\frac{\hbar \omega}{2}(\Op {a^2}-\Op {a^{\dagger}}^2)$ , 
$ \Op {L}=-\frac{\hbar\omega}{2}(\Op {a}^2 +\Op {a^{\dagger}}^2) $, {\text and}

\begin{equation}
Z~~=~~\frac{e^{\frac{\beta\hbar\omega}{2}}}{(e^{\beta\hbar\omega}-1)\sqrt{1-\frac{4\gamma \gamma^{*}}{(e^{\beta\hbar\omega}-1)^{2}}}}~.
\end{equation}

From (\ref{eq:product}), the expectations of $\Op H $ and $  \Op a^2 $ are extracted, leading to
\begin{equation}
\left\langle \Op {H}\right\rangle  ~~ = ~~ \frac{\hbar\omega(e^{2\beta\hbar\omega}-4\gamma \gamma^{*}-1)}{2((e^{\beta \hbar \omega}-1)^{2}-4\gamma \gamma^{*})}
~~and~~
\left\langle \Op {a^2}\right\rangle  ~~ = ~~ \frac{2\gamma^{*}}{(e^{\beta\hbar\omega}-1)^{2}-4\gamma \gamma ^{*}}~.
\label{eq:expeca}
\end{equation}

Equation (\ref{eq:expeca}) can be inverted, leading to
\begin{equation}
\gamma ~~ = ~~ \frac{\frac{\hbar\omega}{2} (\left\langle \Op {L}\right\rangle +i\left\langle \Op {C}\right\rangle )}{\left\langle \Op {L}\right\rangle ^{2}
+\left\langle \Op {C}\right\rangle ^{2}-(\frac{\hbar\omega}{2}-\left\langle \Op {H}\right\rangle )^{2}}
\label{eq:gamainver}
\end{equation}
and the inverse temperature $\beta$:
\begin{equation}
e^{\beta\hbar\omega} ~~ = 
~~ \frac{\left\langle \Op {L}\right\rangle ^{2}+ \left\langle \Op {C}\right\rangle^2-\left\langle \Op {H}\right\rangle ^{2}
+\frac{\hbar^{2}\omega^{2}}{4}}{\left\langle \Op {L}\right\rangle ^{2}
+\left\langle \Op {C}\right\rangle ^{2}-\left(\frac{\hbar\omega}{2}-\left\langle \Op {H}\right\rangle \right)^{2}}
~~.
\label{eq:betainver}
\end{equation}

Equations (\ref{eq:gamainver}) and (\ref{eq:betainver}) relate the state of the system 
$\hat \rho $ (by Equation \ref{eq:product}) to the thermodynamical
observables $\langle \Op {H} \rangle$, $\langle \Op {L} \rangle$, and $\langle \Op {C} \rangle$.

The generalized canonical state of the system Equation (\ref{eq:product}) is equivalent to a squeezed thermal state \cite{kim1989}:
\begin{equation}
\hat \rho = \Op S({\gamma}) \frac{1}{Z}e^{- \beta \Op H} \Op S^{\dagger}(\gamma)~,
\label{eq:sqeez}
\end{equation}
with the squeezing operator
$ \Op S(\gamma)=\exp(\frac{1}{2} (\gamma^* \hat a^2-\gamma \hat a^{\dagger 2}))$. 
This state is an example of a generalized Gibbs state subject to non-commuting constraints \cite{ilievski2015complete,langen2015experimental}. Figure \ref{fig:otto4} shows examples of such states, which all have a Gaussian shape in phase space.

\subsection*{Entropy Balance} 

In thermodynamics, the entropy ${\cal S}$ is a state variable. 
Shannon introduced entropy as a measure
of  missing information required to define a probability distribution $\bf p$ \cite{shanon48}.
The information entropy can be applied to a complete quantum measurement of an observable 
represented by the operator $\Op O$ with possible outcomes $p_j$:
\begin{equation}
{\cal S}_{\Op O}=-k_B \sum_j ~p_j \ln p_j~,
\label{eq:obentropy}
\end{equation}
where $p_j= Tr \{ \Op P_j \hat \rho \}$. 
The projections $\Op P_j$ are defined using the spectral decomposition theorem
$\Op O= \sum_j \lambda_j \Op P_j$, where $\lambda_j$ are the eigenvalues of the operator
$\Op O$. ${\cal S}_{\Op O}$ is then the measure of information gain obtained by the measurement.

The von Neumann entropy \cite{vNeumann} is equivalent to the minimum entropy ${\cal S}_{\Op Y_n}$ 
associated with 
a~complete measurement of the state $\hat \rho$ by the observable $\Op Y_n$, where the set 
of operators $\Op Y_n$ includes all possible non-degenerate operators in Hilbert space. 
The operator that minimizes the entropy commutes with the state $[\hat \rho, \Op Y_{min}]=0$. This leads to a common set of 
projectors of $\Op Y_{min}$ and $\hat \rho$; therefore, ${\cal S}_{VN} = - tr \{ \hat \rho \ln \hat \rho \}$, which is a function of the state only.
Obviously, $ {\cal S}_{VN} \le {\cal S}_{\Op O}$. 
This provides the interpretation that ${\cal S}_{VN}$ is the minimum
information required to completely specify the state $\hat \rho$. 

The primary thermodynamic variable for the heat engine is energy.
The entropy associated with the measurement of energy ${\cal S}_E = {\cal S}_{\Op H}$ in general
differs from the  von Neumann entropy $ {\cal S}_{E} \ge {\cal S}_{VN}$. 
Only when $\hat \rho$ is diagonal in the energy representation---such as in thermal equilibrium (\ref{eq:req})---${\cal S}_{E}= {\cal S}_{VN}$. 

The relative entropy between the state and its diagonal representation in the energy eigenfucntions 
is an alternative measure of coherence \cite{baumgratz2014quantifying}:
 \begin{equation}
D(\hat \rho|| \hat \rho_{ed})=Tr \{ \hat \rho (\ln \hat\rho- \ln \hat \rho_{ed}) \}~,
\label{eq:divergence}
\end{equation}
where $\hat \rho_{ed}$ is the state composed of the energy projections which has the same populations of the energy levels
as state $\hat \rho$.
The conditional distance $D(\hat \rho|| \hat \rho_{ed})$ is equivalent to the difference between the energy entropy ${\cal S}_E$
and the von Neumann entropy ${\cal S}_{VN}$: $D(\hat \rho|| \hat \rho_{ed})={\cal S}_E-{\cal S}_{VN} \ge 0$.

The von Neumann entropy is invariant under unitary evolution \cite{alhassid}. 
This is the result of the property of unitary transformations, 
where the set of eigenvalues of $\hat \rho' =\Op U \hat \rho \Op U^{\dagger}$ 
is equal to the set of eigenvalues of  $\hat \rho$. Since the von Neumann entropy ${\cal S}_{VN}$ 
is a functional of the eigenvalues of $\hat \rho$, it becomes invariant
to  any unitary transformation.

When the  unitary transformation  is generated by  members of the Lie algebra, the Casimir is invariant.
The von Neumann entropy of the generalized Gibbs state (\ref{eq:product}) is a function of the Casimir $\langle \Op G \rangle $ \cite{k243}
so that in this case it also becomes constant:
\begin{eqnarray}
{\cal S}_{VN}=\ln \left(\sqrt{ \langle \Op G \rangle -\frac{1}{4}} \right)+\sqrt{\langle \Op G \rangle} asinh \left( \frac{\sqrt{\langle \Op G \rangle}}{{ \langle \Op G \rangle -\frac{1}{4}}}\right)~.
\label{eq:vonneumann}
\end{eqnarray}

An alternative expression for the ${\cal S}_{VN}$ entropy is calculated from  the covariance matrix of Gaussian canonical states \cite{isar1999purity,serafini2003symplectic,brown2016passivity}:
\begin{eqnarray}
{\cal S}_{VN}=\frac{\nu+1}{2} \ln (\frac{\nu+1}{2})-\frac{\nu-1}{2} \ln (\frac{\nu-1}{2})~,
\label{eq:vonneumann2}
\end{eqnarray}
where $\nu =\frac{2}{\hbar} \sqrt{\sigma}$, 
\begin{eqnarray}
\sigma =
\left|
\begin{array}{cc}
\sigma_{pp}&\sigma_{pq}\\
\sigma_{qp}&\sigma_{qq}
\end{array}
\right|~,
\nonumber
\end{eqnarray}
and $\sigma_{ij}$ is the covariance.

The energy entropy ${\cal S}_E$ of the oscillator (not in equilibrium) 
is found to be equivalent to the entropy of
an oscillator in thermal equilibrium with the same energy expectation value:
\begin{equation}
{\cal S}_E~~=~~\frac{1}{\hbar\omega}\left(\left\langle {\Op {H}}\right\rangle 
+\frac{\hbar\omega}{2}\right)
\ln\left(\frac{\left\langle \mathbf{\hat{H}}\right\rangle + \frac{\hbar \omega}{2}}
{\left\langle \mathbf{\hat{H}}\right\rangle -\frac{\hbar\omega}{2}}\right)
-\ln\left(\frac{\hbar\omega}{\left\langle \mathbf{\hat{H}}\right\rangle -\frac{\hbar\omega}{2}}\right)~.
\label{eq:enentrop1}
\end{equation}

${\cal S}_E $ in (\ref{eq:enentrop1}) is completely determined by the energy expectation 
$E=\left\langle \mathbf{\Op {H}}\right\rangle$. As an extreme example, for~a~squeezed pure state,
${\cal S}_{VN}=0$ and ${\cal S}_E \ge 0$. 

In a macroscopic working medium, the internal temperature
can be defined from the entropy and energy variables
$1/T_{int} = \left(\frac{\partial {\cal S}}{\partial E}\right)_V$ at constant volume.
For the quantum Otto cycle, ${\cal S}_E$ is used to define the inverse internal temperature
$1/T_{int} = \left(\frac{\partial {\cal S}_E}{\partial E}\right)_{\omega}$. 
$T_{int}$ is a generalized  temperature appropriate for non  equilibrium density 
operators $\hat \rho$. Using this definition, the internal temperature $T_{int}$ of the oscillator 
working medium can be calculated implicitly from the energy expectation:
\begin{equation}
E=\frac{1}{2}\hbar\omega\coth\left(\frac{\hbar\omega}{2 k_B T_{int}}\right)~,
\label{eq:inittemp}
\end{equation}
which is identical to the equilibrium relation between temperature and energy in the harmonic oscillator.
This temperature defines the work required to generate the coherence: ${\cal W}_c = k_B T_{int}({\cal S}_E- {\cal S}_{VN}) $~\cite{plastina2014}.

\section{The Dynamics of the Quantum Otto Cycle}
\label{sec:otodynamics}

A quantum heat engine is a dynamical system subject to the tradeoff between efficiency and power. 
The dynamics of the reciprocating Otto cycle can be partitioned to the four strokes and later combined to  generate the full cycle.
Each of the segments influences the final performance: power extraction or refrigeration.
The performance of the cycle can be optimized with respect to efficiency and power.
Each segment can be optimized separately, and finally a global optimization is performed.
The first step is to describe the dynamics of each segment in detail.

\subsection{Heisenberg Dynamics of Thermalisation on the {{Isochores}}}

The task of the {\em isochores} is to extract and reject heat from thermal reservoirs. 
The dynamics of the working medium is dominated by an approach to thermal equilibrium. In the Otto cycle, the 
Hamiltonian $\Op H$ is constant ($\omega =\omega_{h/c}$ is constant). 
The Heisenberg equations of motion generating the dynamics for an operator $\Op X$ become:
\begin{eqnarray}
\frac{d}{dt}{\Op X} 
=\frac{i}{\hbar}[\Op H,\Op X]+
k_{\downarrow}(\Op a^{\dagger}\Op X \Op {a}-\frac{1}{2}\{\Op {a}^{\dagger}\Op {a},\Op X\})+
k_{\uparrow}(\Op{a}\Op X \Op {a}^{\dagger}-\frac{1}{2}\{\Op{a}\Op{a}^{\dagger},\Op X\})~.
\label{eq:hieniso}
\end{eqnarray}

Equation (\ref{eq:hieniso}) is the analogue of (\ref{eq:lvn2}) and (\ref{eq:dissipative}) 
in the Schr\"odinger frame.

For the dynamical set of observables, the equations of motion become:
\begin{eqnarray}
\frac{d}{dt}\left(\begin{array}{c}
\Op H\\
\Op L\\
\Op C\\
\Op I
\end{array}\right)(t)=
\left(\begin{array}{cccc}
-\Gamma & 0 & 0 & \Gamma \langle \Op H \rangle_{eq}\\
0 & -\Gamma & -2\omega & 0 \\
0 & 2 \omega & -\Gamma & 0\\
0&0&0&0
\end{array}\right)
\left(\begin{array}{c}
\Op H\\
\Op L\\
\Op C\\
\Op I
\end{array}\right)(t)~,
\label{eq:motion}
\end{eqnarray}
where $\Gamma=k_{\downarrow}-k_{\uparrow}$ is the heat conductance and 
$k_{\uparrow}/k_{\downarrow}= e^{-\hbar\omega/k_B T}$ obeys detailed balance where
$\omega=\omega_{h/c}$ and $T=T_{h/c}$ 
are  defined for the hot or cold bath, respectively. From (\ref{eq:firstlaw}), the heat current can be identified as:
\begin{eqnarray} 
\dot {\cal Q}=-\Gamma ( \langle \Op H \rangle -\langle \Op H \rangle_{eq})~~=~~\Gamma \frac{\hbar \omega}{2}
\left( \coth ( \frac{\hbar \omega}{2 k_B T_B})-\coth ( \frac{\hbar \omega}{2 k_B T_{int}}) \right)~,
\label{eq:hettrans}
\end{eqnarray}
where $T_B$ is the bath temperature.
In the high temperature limit, the heat transport law becomes Newtonian \cite{curzon75}: 
$ \dot {\cal Q}=\Gamma (k_B T_B -K_B T_{int})$.

The solution of {\em isochore} dynamics (\ref{eq:motion}) generates the propagator defined on the vector space of the observables
$\Op H, \Op L,\Op C,\Op I$:
The propagator on the {\em isochore} has the form \cite{k243,insinga2016thermodynamical}:
\begin{eqnarray}
{\cal U}_{h/c} = 
\left(
\begin{array}{cccc}
R&0&0&H_{eq}(1-R)\\
0&Rc&-R s&0\\
0&Rs&  Rc&0\\
0&0&0&1
\end{array}
\right)~,
\end{eqnarray}
where $R=e^{-\Gamma t}$. $c=\cos (2 \omega t)$, $s=\sin (2 \omega t)$,
and $H_{eq}= \frac{\hbar \omega}{e^{\frac{\hbar \omega}{k T}} -1}$.
It is important to note that the propagator on the {\em isochores} does not generate coherence from energy  $\Op H$. The
coherence Equation (\ref{eq:coherence})  is a function of the expectations of $\Op L,\Op C$, which are not coupled to $\Op H$.

\subsection{The Dynamics on the  Adiabats and Quantum Friction}
\label{subsec:adidy}

The dynamics on the {\em adiabats} is generated by a time-dependent Hamiltonian.
The task is to change the energy scale of the working medium from one bath to the other.
The oscillator frequency changes from $\omega_h$ to $\omega_c$ on the power
expansion segment and from $\omega_c$ to $\omega_h$ on the compression segment. 
The~Hamiltonian---which is explicitly time-dependent---does not commute with itself at different times $[\Op H (t),\Op H(t')] \neq 0$.
As a result, coherence is generated with an extra cost in energy.

The Heisenberg equations of motion (\ref{eq:heisenberg}) for the dynamical set of operators are expressed \mbox{as \cite{k243,insinga2016thermodynamical}}:
\begin{equation}
\frac{d}{dt}\left(\begin{array}{c}
\Op H\\
\Op L\\
\Op C\\
\Op I
\end{array}\right)(t)=\omega(t)
\left(\begin{array}{cccc}
\mu & -\mu & 0&0\\
-\mu & \mu & -2&0\\
0 & 2 & \mu&0\\
0&0&0&0
\end{array}\right)
\left(\begin{array}{c}
\Op H\\
\Op L\\
\Op C\\
\Op I
\end{array}\right)(t)~,
\label{eq:adiabatdy}
\end{equation}
where $\mu=\frac{\dot \omega}{\omega^2}$ is a dimensionless adiabatic parameter.
In general, all operators in (\ref{eq:adiabatdy}) are dynamically coupled. This coupling
is characterized by the non-adiabatic parameter $\mu$.
When $\mu \rightarrow 0$, the energy decouples from the coherence and the cycle can be characterized 
by $p_n$---the probability of occupation of energy level $\epsilon_n$.

Power is obtained from the first-law (\ref{eq:firstlaw}) as:
\begin{equation}
{\cal P} = \mu \omega \left( \langle \Op H \rangle -\langle  \Op L \rangle \right)~.
\label{eq:pl}
\end{equation}

Power on the {\em adiabats} (\ref{eq:pl}) can be decomposed to the ``useful'' external power
${\cal P}_{ex}= \mu \omega \langle  \Op H \rangle $ and to the power invested to counter friction
${\cal P}_{f}= -\mu \omega \langle  \Op L \rangle $ if $\langle  \Op L \rangle > 0$. 
Under adiabatic conditions $\mu \rightarrow 0$, $\langle \Op L \rangle =0 $, since no coherence is generated; therefore, ${\cal P}_{f}=0$.
Generating coherence consumes power when the initial state is diagonal in energy $[\hat \rho, \Op H]=0$~\cite{zagoskin2012squeezing,brandner2017universal}.

Insight on the adiabatic dynamics can be obtained from the closed-form solution of the dynamics
when the non-adiabatic parameter $\mu=\frac{\dot \omega}{\omega^2}$ is constant. 
This leads to the explicit time dependence of the control frequency $\omega$: 
$\omega(t)=\frac{\omega(0)}{1 - \mu \omega(0)t}$.
Under these conditions, the matrix in Equation (\ref{eq:adiabatdy}) becomes stationary. 
This allows a closed-form solution to be obtained by diagonalizing the matrix. 
Under these conditions, the adiabatic propagator ${\cal U}_a$ has the form:
\begin{eqnarray}
{\cal U}_a = \frac{\omega(t)}{\omega(0)}\frac{1}{\Omega^2}
\left(
\begin{array}{cccc}
4-\mu^2 c &-\mu \Omega s &-2 \mu(c-1)&0\\
-\mu \Omega s&\Omega^2 c&-2 \Omega s&0\\
 2 \mu(c-1)& 2 \Omega s&4c- \mu^2&0\\
0&0&0&1
\end{array}
\right)~,
\label{eq:adprop}
\end{eqnarray}
where $\Omega=\sqrt{4-\mu^2}$ and $c=\cos (\Omega \theta(t))$, $s=\sin (\Omega \theta(t))$, and $\theta(t)=-\frac{1}{\mu}\log (\frac{\omega(t)}{\omega(0)}) $
\cite{k243}. For $|\mu| < 2$, the solutions are oscillatory. For $|\mu| > 2$, the $\sin$ and $\cos$ functions become $\sinh$ and $\cosh$.
More on the transition point from damped to over-damped  dynamics is presented in Section \ref{subsec:ep}.

The difference between the expansion {\em adiabats} ${\cal U}_{hc}$ 
and compression {\em adiabats} ${\cal U}_{hc}$ is in the sign of $\mu$ and
the ratio $\omega(t)/\omega(0)$. The propagator ${\cal U}_a$ can be viewed as a product of a changing energy scale by the factor 
$ \frac{\omega(t)}{\omega(0)}$ and a propagation in a moving frame generated by a constant matrix.
The fraction of additional work on the {\em adiabats} with respect to the {\em adiabatic} 
solution is causing extra energy invested in the woking medium, which  is defined as: 
\begin{equation}
\delta_f=\frac{\omega_i}{\omega_f} ({\cal U}_a (1,1) -\frac{\omega_f}{\omega_i})~.
\label{eq:deld}
\end{equation}

For the case of $\mu$ constant:
\begin{equation}
\delta_f = \frac{ 2 \mu^2 \sin(\frac{\theta \Omega}{2})^2}{ 4-\mu^2}
\label{eq:del}
\end{equation}
and $\delta_f \ge 0$.

A different approach to the deviation from adiabatic behaviour has been based on a general propagator for Gaussian wavefunctions of the form \cite{deffner2010quantum}:
\begin{equation}
\psi(x,t)=\exp\left( \frac{i}{2 \hbar} ( a(t) x^2 +b(t) x + c(t) )\right)~.
\label{eq:lutz}
\end{equation}
$a(t)$ can be mapped to a time-dependent classical harmonic oscillator: $a(t) = M \dot X /X $, where:
\begin{equation}
m \frac{d^2}{dt^2} X + \omega(t)^2 X =0~.
\label{eq:xx}
\end{equation}

The local adiabatic parameter is defined as:
\begin{equation}
Q^*(t) = \frac{1}{2 \omega_i \omega_f}\left(\omega_i^2 (\omega_f^2 X(t)^2+\dot X(t)^2) + (\omega_f^2  Y(t)^2 + \dot Y(t)^2) \right)~,
\label{eq:Q}
\end{equation}
where $X(t) $ and $Y(t)$ are the solution of Equation (\ref{eq:xx}) with the boundary conditions $X(0)=0$, $\dot X(0)=1$ and $Y(0)=1$, $\dot Y(0)=0$ for a constant frequency $Q^*=1$. In general, the expectation value of the energy at the end of the {\em adiabats} becomes:
\begin{equation}
\langle \Op H \rangle_f = \frac{\omega_f}{\omega_i} Q^* \langle \Op H \rangle_i~,
\end{equation} 
where $i/f$ correspond to the beginning and end of the stroke. In general,  $Q^*(t)$ can be obtained directly from the solution of Equation (\ref{eq:xx}).
$Q^*(t)$ is related to $\delta_f$ by: $Q^*(t) = 1 +\delta_f$. For the case of $\mu$, constant  $Q^*(t)$ can be obtained from Equation (\ref{eq:del}).
In addition, $Q^*(t)$ can be obtained by the solution of the Ermakov equation (Equation (\ref{eq:ermakov})) \cite{beau2016scaling}.

The general dynamics described in Equation (\ref{eq:adprop}) mixes energy and coherence. As can be inferred from the Casimir Equation (\ref{eq:casimir}),
generating coherence costs energy. This extra cost gets dissipated on the {\em isochores}, and is termed {\em quantum friction} \cite{k190,plastina2014}. 
The energy cost scales as $\mu^2$; therefore, slow operation (i.e., $|\mu| \ll 1$) will eliminate this cost. The drawback is large cycle times
and low power. 
Further analysis of Equation (\ref{eq:adprop}) shows a surprising result. Coherence can be generated and consumed, resulting in periodic solutions in which the propagator becomes diagonal. As a result, mixing between energy and coherence is eliminated. These solutions appear
when $\cos (\Omega \theta(\tau_a))=1$, where $\tau_a$ is the expansion or compression stroke time allocation. 
These periodic solutions can be characterised by a~quantization relation \cite{k243}: 
\begin{equation}
\mu^*= - \frac{2 \log ({\cal C})}{\sqrt{4\pi^2 l^2+\log({\cal C})^2}}~,
\label{eq:mucrit}
\end{equation}
where ${\cal C}=\frac{\omega_c}{\omega_h}$ is the engine's compression ratio, and $l$ the quantization number $l=1,2,3,...$,
accompanied by the  time allocation $\tau_{hc}^*$:
\begin{equation}
\tau_{hc}^*=\frac{1- {\cal C}}{\mu^* \omega_c}~.
\label{eq:tperiodic}
\end{equation}

A frictionless solution with the shortest time is obtained for $l=1$, and it scales as 
$\tau_{hc} \propto 1/\omega_c$. 

This observation raises the question:
are there additional frictionless solutions in finite time? What is the shortest time that can achieve this goal?

The general solution of the dynamics depends on an explicit dependence of $\omega(t)$ on time. 
$\omega(t)$~can be used as a control function to optimise the performance, 
obtaining a state $\hat \rho$  diagonal in energy at the interface with the {\em isochores}.
Such a solution will generate a frictionless performance. 
Operating at effective adiabatic conditions has been termed  {\em shortcut to adiabaticity} \cite{chen2010fast,torrontegui2013shortcuts,chen2010,muga2010transitionless,muga2016}.

The search for frictionless solutions has led to two main directions. The first is based on a time-dependent invariant operator $\Op I(t)$ \cite{muga2016}:
\begin{equation}
\frac{d}{dt} \Op I(t) = \frac{\partial}{\partial t} I(t) + \frac{i}{\hbar} [ \Op H(t), \Op I(t)] = 0~.
\label{eq:invar}
\end{equation}

For the harmonic oscillator,  the invariant is \cite{chen2010fast}:
\begin{equation}
\Op I(t) = \frac{1}{2}\left ( \frac{1}{b^2}\Op Q^2 m \omega_0^2 + \frac{1}{m} \Op \pi^2 \right)~,
\end{equation}
where $\Op \pi  = b \Op P - m \dot b \Op Q$. The invariant must satisfy $[\Op H, \Op I]=0$ for the initial  $\tau_i$ and final $\tau_f$ times,
then at these times the eigenstates of the invariant at the initial and final time of the {\em adiabat}
are identical to those of the Hamiltonian \cite{torrontegui2011fast,chen2011lewis}. 
This is obtained if $ b(0)=1$ and $\dot b(0)=0$, $\ddot b(0)=0$, as well as $b(\tau_f)=\sqrt{\omega_0/\omega_f}$ and $\dot b(\tau_f)=0$,
$\ddot b(\tau_f)=0$. In addition, the function $b(t)$ satisfies the Ermakov~equation:
\begin{equation}
\ddot b + \omega(t)^2 b = \omega_0^2/b^3~.
\label{eq:ermakov}
\end{equation}

The instantaneous frequency becomes $\omega(t) = \omega_0 /b^2$ for $\ddot b(0)=0$. 
There are many solutions to the Ermakov equation, and additional constraints
must be added---for example, that the frequency is at all times real and positive. These equations can be used to search for fast frictionless solutions.
To obtain a minimal time $\tau_a$, some constrains have to be imposed. For example, limiting the average energy stored in the oscillator. In this case,
$\tau_{hc}$ scales as $\tau_{hc}^* \propto 1/\sqrt{\omega_c}$ \cite{chen2010}.
Other constraints on $\omega(t)$ have been explored. For example, the use of imaginary frequency corresponding to
an inverted harmonic potential. These schemes allow faster times on the {\em adiabat} \cite{chen2010,k269}. If the peak energy is constrained, 
$\tau_{hc}$ scales logarithmically with $1/\omega_c$; however, if the average energy is constrained, then the scaling becomes $\tau_{hc}^* \propto 1/\sqrt{\omega_c}$.

The second approach to obtain frictionless solutions is based on optimal control theory: finding the fastest frictionless solution where the control function is $\omega(t)$ \cite{k243,k242,k269,salamon2012optimal,hoffmann2013optimal,boldt2016fastest,PhysRevE.94.022141}. Optimal control theory reveals that the problem of minimizing time is linear in the control, 
which is proportional to $\omega(t)$ \cite{k243}. As a result, the optimal control solution depends on the constraints 
$\omega_{max}$ and $\omega_{min}$.
If these are set as $\omega_{max}=\omega_h$ and $\omega_{min}=\omega_c$, then the optimal time scales as 
$\tau_a^* \propto \frac{1}{(\sqrt{\omega_c}\sqrt{\omega_c})}$. Other constraints will lead to faster times, but their energetic cost will diverge. 
This scaling is consistent considering the cost of the counter-adiabatic terms in frictionless solutions
leading to the same scaling \cite{campbell2016trade}.

The optimal solution can be understood using a geometrical description \cite{salamon2012optimal}. 
The derivative of the change of $\langle \Op Q^2 \rangle$
with respect to the change in $\langle \Op P^2 \rangle$ becomes:
\begin{equation}
\frac{d \langle \Op Q^2\rangle }{ d \langle  \Op P^2 \rangle } = - \omega^2(t) \equiv v~.
\end{equation}

The time allocated to the change $\tau$ becomes:
\begin{equation}
\tau= \int_{ \langle \Op P^2 \rangle_i}^{ \langle \Op P^2 \rangle_f} \frac{ d \langle \Op P^2 \rangle }
{\sqrt{ \langle \Op P^2 \rangle \langle \Op Q^2 \rangle- \langle \Op G \rangle}}~,
\end{equation}
where $ \Op  G$ is the Casimir defined in Equation (\ref{eq:casimir}). In addition, the control $v$ is constrained by $\omega_c^2 \le v \le \omega_h^2$.
The initial and final  $\langle \Op P^2 \rangle_{i/f}= m E_{i/f}$, since the initial and final $\langle \Op L \rangle$ and $\langle \Op C \rangle $ are zero.
The~minimum time is obtained by maximizing the product $\langle \Op P^2 \rangle \langle \Op Q^2 \rangle$ along the trajectory.
The minimum time optimization leads to a bang-bang solution where the frequency is switched instantly from $\omega_h$ to $\omega_c$, as in
the sudden limit Equation (\ref{eq:sudden})  is followed by a waiting period
then switched back to $\omega_h$ until the target is reached and switched finally to $\omega_c$. 
The relation between the geometric optimization and the Ermakov equation of the shortcuts to adiabaticity has been obtained 
based on the geometrical optimization \cite{stefanatos2010,stefanatos2016minimum}.

To summarize, frictionless solutions can be obtained in finite time. As a result, the engine can be completely described by the population of the energy eigenvalues or  for the harmonic working medium by the 
expectation value of number operator $\Op N$.
Employing reasonable constraints on the control function  $\omega (t)$ results in the minimum time 
$\tau_a^*$ scalling as $O(\frac{1}{\sqrt{\omega_c}\sqrt{\omega_h}})$. 

\subsection{The Influence of Noise on the Adiabats}
\label{subsec:noise}

The frictionless {\em adiabat} requires a very accurate protocol of $\omega (t)$ as a function of time.
For any realistic devices, such a protocol will be subject to fluctuations in the external control. 
The controllers are subject  to noise, which will induce friction-like behaviour. Can this additional friction be minimized?
Insight on the effects of noise on the performance of the Otto cycle can be obtained by analysing a~simple model based on 
the frictionless protocol with constant $\mu$
~\cite{k283}.
The obvious source of external noise  is induced by fluctuations in the control frequency $\omega (t)$. 
This noise is equivalent to Markovian random fluctuations in the frequency of the harmonic oscillator. 
These errors are modelled by a~Gaussian white noise. 
The dissipative Lindbland term generating such noise has the form \cite{gorini76,breuer02}: 
\begin{equation}
\mathcal{L}_{N_{a}}(\bold{\hat A})=-\gamma_a\omega^2[\bold{\hat B},[\bold{\hat B},\bold{\hat A}]],
\end{equation}
where $\bold{\hat B}=m\omega\bold{\hat Q}^2/(2\hbar)$.

The influence of the amplitude noise generated by  
$\mathcal{L}_{N_{a}}(\bold{\hat A})=-\gamma_a\omega^2[\bold{\hat B},[\bold{\hat B},\bold{\hat A}]]$ is obtained by 
approximating the propagator by the product form $\mathcal{U}_{hc}=\mathcal{U}_{a} \mathcal{U}_{an}$. 
The equations of motion for the amplitude noise $\mathcal{U}_{an}$ are obtained from the interaction picture in Liouville space:
\begin{eqnarray}
\nonumber
\frac{d}{\omega dt}\mathcal{U}_{an}(t)&=&\mathcal{U}_a(-t)
\mathcal{N}_a(t)
\mathcal{U}_a(t)\mathcal{U}_{an}(t)
\\
&=&\mathcal{W}_{an}(t)\mathcal{U}_{an}(t)~,
\label{eq:fffff}
\end{eqnarray}
where ${\cal W}_{an}$ is the interaction propagator in Liouville space \cite{k283} and 
${\cal U}_a$ is the {\em adiabatic} propagator, Equation (\ref{eq:adprop}).
A closed-form solution is obtained  in the frictionless limit $\mu\rightarrow 0$ when  $\mathcal{W}_{a}$ is expanded up to zero order in $\mu$:
\begin{eqnarray}
\mathcal{W}_a(t)\approx{\gamma_a\omega_0}
\left(\begin{array}{cccc} 
1  & -c & s&0 \\
c   &   -c^2 & cs&0 \\
-s &    cs & -s^2&0\\
0&0&0&0\\
\end{array} \right)~,
\end{eqnarray}
where $s =\sin (\Omega \Theta)$ $c=\cos (\Omega \Theta)$.
The Magnus expansion \cite{blanes} is employed to obtain the $l$ period propagator $\mathcal{U}_{3a}(X=2l\pi)$, where the periods are of the {\em adiabatic} propagator ${\cal U}_a$ of Equation (\ref{eq:adprop}): 
\begin{equation}
\mathcal{U}_{3a}(X=2l \pi)\approx e^{B_1 + B_2 + \dotsc}~,
\end{equation}
where $B_1=\int_{0}^{2n\pi} dX\mathcal{W}_{a}(X)$, $B_2=\frac{1}{2}\int_{0}^{2n\pi}\int_{0}^{X} dX dX' [\mathcal{W}_{a}(X),\mathcal{W}_{a}(X')]$, and so on. 
The first-order Magnus term leads to the propagator
\begin{eqnarray}
\nonumber
\label{U3a}
\mathcal{U}_{an}(X&=&2l\pi)_{B_{1}}
=
\left(\begin{array}{cccc} 
e^{\gamma_a {\cal F} / \mu}  &  0 & 0&0\\
0   &   e^{-\gamma_a {\cal F}/(2 \mu)}   & 0&0 \\
0&    0 & e^{-\gamma_a {\cal F} /(2 \mu)}  &0\\
0&0&0&1
\end{array} \right)~,
\end{eqnarray}
where ${\cal F} = \left(\frac{ 16 \omega_0}{(-16 + 3 \mu^2)}\right)\left(e^{2 \pi l \mu \Omega}-1\right)$. For large $l$, in
~Equation (\ref{eq:mucrit}) the limit from hot to cold simplifies to:
${\cal F} = (\omega_h-\omega_c)$.
The solution of Equation (\ref{eq:fffff}) shows that the fraction of work against friction $\delta_f$ will diverge when $l \rightarrow \infty$ or $\mu \rightarrow 0$,
nulling the adiabatic solution for even a very small $\gamma_a$.
The best way to eliminate amplitude noise is to choose the shortest frictionless protocol. 
Nevertheless, some friction-like behaviors will occur.

Next, phase noise is considered. It occurs due to errors in the piecewise process used for controlling the scheduling of $\omega$ in time. 
For such a procedure, random errors are expected in the duration of the time intervals.
These errors are modeled by a Gaussian white noise. Mathematically, the process
is equivalent to a dephasing process on the {\it adiabats} \cite{k215}. 
The dissipative operator $\mathcal{L}_N$ has the form given by \cite{gorini76,breuer02}:
\beq
\mathcal{L}_{N_{p}}(\bold{\hat A})=-\frac{\gamma_p}{\hbar^2}[\bold{\hat H},[\bold{\hat H},\bold{\hat A}]]~.
\eeq

In this case, the interaction picture for the phase noise ${\cal U}_{p}$ becomes
\begin{eqnarray}
\nonumber
\frac{d}{\omega dt}{\cal U}_{pn}(t)&=&{\cal U}_a(-t)
{\cal N}_p(t)
{\cal U}_a(t)\mathcal{U}_{pn}(t)
~=~{\cal W}_p(t){\cal U}_{pn}(t)~,
\end{eqnarray}
which at first order in $\mu$ can be approximated as 
\begin{eqnarray}
\nonumber
{\cal W}_p(t)~\approx~
2\gamma_p\omega_0&\times&
\left(\begin{array}{cccc} 
0  & \mu s & \mu (1-c) &0\\
\mu s   &  -(2+\mu X) & 0&0 \\
\mu (c-1)&    0 & -(2+\mu X)&0\\
0&0&0&0
\end{array} \right).
\end{eqnarray}

Again, using the Magnus expansion for one period of $X$ leads to 
{\fontsize{9}{9}\selectfont
\begin{eqnarray}
\nonumber
{\cal U}_{3p}(X=2\pi)_{B_{1}}=
\left(\begin{array}{cccc} 
1  &  0 & (1-e^{8\pi\gamma_p\omega_0})\mu/2&0\\
0   &   e^{8\pi\gamma_p\omega_0}(1-4\pi^2\mu\gamma_p\omega_0)   & 0 &0\\
 (-1+e^{8\pi\gamma_p\omega_0})\mu/2&    0 & e^{8\pi\gamma_p\omega_0}(1-4\pi^2\mu\gamma_p\omega_0) &0\\
 0&0&0&1
\end{array} \right)~.
\end{eqnarray}}

At first order in $\mu$, this evolution operator maintains $\delta_f(1)=0$, so the frictionless case holds. 
The~second-order Magnus term leads to the noise correction
\begin{eqnarray}
\label{U3p}
{\cal U}_{3p}(X=2\pi)_{B_{2}}=
\left(\begin{array}{cccc} 
\cosh \beta & -\sinh \beta   & 0&0\\
-\sinh \beta      &   \cosh \beta     & 0 &0\\
0 &    0 & 1&0\\
0&0&0&1
\end{array} \right)~,
\end{eqnarray}
where $\beta= \frac{16 \omega_0^2 \gamma_p^2}{4 +3 \mu^2}\left(e^{2 \pi l \mu \Omega}-1\right)$. 
In the limit of $l \rightarrow \infty$, $\beta = 4 \gamma^2(\omega_h^2-\omega_c^2)$.
The propagator ${\cal U}_{3p}(X=2\pi)_{B_{2}}$ mixes energy and coherence, even at the limit 
$\mu \rightarrow 0$ and $\tau_a \rightarrow \infty$, where one would expect frictionless solutions.

We can characterize the fraction of 
additional energy generated by a parameter $\delta$. Asymptotically
for amplitude noise: $\delta_a = e^{\gamma_a {\cal F}/\mu} > 0$,  and for phase noise $\delta_f(1)=e^{\gamma_p {\cal F}\mu} -1 \approx 0$,
and the second-order correction $\delta_f(2)= \cosh(\beta) -1 >0 $.
Imperfect control on the {\em adiabats} will always lead to $\delta_f >0$ and additional work invested in friction.

\subsection{The Sudden Limit}

The limit of vanishing time on the {\em adiabats} $\tau_a \ll 1/\omega_c$ leads to the sudden propagator; therefore, $\mu \rightarrow \pm \infty$.
Such dynamics is termed {\em sudden quench}.
The propagator ${\cal U}_a$ has an explicit expression:
\begin{eqnarray}
{\cal U}_{a} =\left(
\begin{array}{cccc}
\frac{1}{2}(1+\alpha)&\frac{1}{2}(1-\alpha)&0&0\\
\frac{1}{2}(1-\alpha)&\frac{1}{2}(1+\alpha)&0&0\\
0&0&1&0\\
0&0&0&1
\end{array}
\right)~,
\label{eq:sudden}
\end{eqnarray}
where $\alpha = (\frac{\omega_f}{\omega_i})^2$ is related to the compression ratio $\alpha_{ch}={\cal C}^2$ and 
$\alpha_{hc}={\cal C}^{-2}$. The propagator mixes $\Op H$ and $\Op L$ when the compression ratio deviates from 1.
As a result coherence is generated.
The sudden propagator is an integral part of the frictionless bang-bang solutions \cite{k243,salamon2012optimal}.
Equation (\ref{eq:sudden}) can be employed as part of a bang-bang {\em adiabat} or as part of a complete sudden cycle.

\subsection{Effects of an Exceptional Point on the Dynamics on the Adiabat}
\label{subsec:ep}

Exceptional points (EPs) are degeneracies of non-Hermitian
dynamics \cite{kato,heis} associated with the coalescence
of two or more eigenstates. The studies of
EPs have substantially grown due to the observation of  (space-time reflection symmetry) PT symmetric
Hamiltonians \cite{bender2007making}. These Hamiltonians have a real spectrum,
which~becomes complex at the EP. The main effect of
EPs (of any order) on the dynamics of PT-symmetric
systems is the sudden transition from a real spectrum to
a complex energy \mbox{spectrum~\cite{nimrod08,k301}}.

The adiabatic strokes are generated by a time-dependent Hamiltonian Equation (\ref{eq:hamils}). 
We therefore expect 
the propagator ${\cal U}_a$ to be unitary, resulting in eigenvalues with the property $|u_j|=1$. These properties are only true
for a compact Hilbert space. 
We find surprising exceptions for the non-compact harmonic oscillator with an~infinite number of energy levels.
We can remove the trivial scaling $\frac{\omega(t)}{\omega(0)}$ in Equation (\ref{eq:adiabatdy}) which originates from the diagonal part.
The propagator can be written as ${\cal U}_a = {\cal U}_0 {\cal U}_1$, where ${\cal U}_0 = \frac{\omega(t)}{\omega(0)} {\cal I}$ is a rescaling of the energy unit.
The equation of motion for ${\cal U}_1$ becomes:
\begin{equation}
\frac{d}{d\theta}{\cal U}_1(\theta)=
\left(\begin{array}{cccc}
0& -\mu & 0\\
-\mu & 0 & -2\\
0 & 2 & 0\\
\end{array}\right)
{\cal U}_1(\theta)~,
\label{eq:adiabatdy-2}
\end{equation}
where the trivial propagation of the identity is emitted and
the time is rescaled $\theta (t) =\int^t \omega(t') dt'$.
Diagonalising Equation (\ref{eq:adiabatdy-2}) for constant $\mu$,
we can identify three eigenvalues: $\lambda_1=0$ and 
$\lambda_{23} = \pm i \sqrt{ 4 -\mu^2}$. For $\mu \le 2$, as expected, Equation (\ref{eq:adiabatdy-2}) generates a unitary propagator.
The three eigenvalues become degenerate when $\mu=2$, and become real for  $\mu \ge 2$ $\lambda_{23} $   \cite{k282}. 
This is possible because  the generator Equation (\ref{eq:adiabatdy-2}) is non-Hermitian. 
At the exceptional point, the matrix in Equation (\ref{eq:adiabatdy-2}) has a single eigenvector corresponding to $\lambda_1=\lambda_{23}$,
which is self-orthogonal. To show this property, it is necessary to multiply the right and left eigenvectors of the non-symmetric matrix  
at the EP.
Their~product is equal to zero, showing that the eigenvector is self-orthogonal \cite{nimrod-book}.
The propagator Equation (\ref{eq:adprop}) changes character at the EP; $\Omega = \sqrt{4-\mu^2}$ changes from a real to  an imaginary number.
As a result, the dynamics at the EP changes from oscillatory to exponential \cite{k282}.

This effect can also be observed in the classical parametric oscillator Equation (\ref{eq:xx}).
By changing the time variable $\frac{d}{dt}=\omega(t) \frac{d}{d \tau}$, and for constant $\mu$,
~the equation of motion becomes
\begin{equation}
\left( \frac{d^2}{d \tau^2} + \mu \frac{d}{d \tau} +1 \right) X(\tau)=0~,
\label{eq:mumu1}
\end{equation}
which is the well-known equation of motion of a
damped harmonic oscillator. Note that the original model (given by Equation (\ref{eq:xx})) does not involve dissipation, and a priori one
would not expect the appearance of an EP. The rescaling of
the time coordinate allows us to identify an EP at $|\mu|=2$,
corresponding to the transition between an underdamped
and an over-damped oscillator \cite{moiseyev2013}. 

Exceptional points are also expected in the eigenvalues and eigenvectors of the total propagator ${\cal U}_{cyc}$, which posses complex eigenvalues. 
Such points will indicate a drastic change in the cycle~performance.

\section{Closing the Cycle}

Periodically combining  the four propagators leads to the cycle propagator. Depending on the choice of parameters,
we get either an engine cycle where heat flow is converted to power:
\begin{equation}
 {\cal U}_{cyc}^e={\cal U}_{ch}{\cal U}_c {\cal U}_{hc}{\cal U}_h~~~{\text {where}}~~\frac{ \omega_c}{\omega_h} > \frac{T_c}{T_h}~,
 \label{eq:cycengin}
 \end{equation}
 or a refrigerator cycle where power drives a heat current from the cold to the hot bath:
 \begin{equation}
 {\cal U}_{cyc}^r={\cal U}_{ch}{\cal U}_c {\cal U}_{hc}{\cal U}_h~~~{\text {where}}~~\frac{ \omega_c}{\omega_h} < \frac{T_c}{T_h}~.
 \label{eq:cycref}
 \end{equation}
 
In both cases, $\omega_h > \omega_c$.
Frictionless cycles are either refrigerators or engines. Friction adds another possibility.
When the internal friction dominates both, the engine cycle and the refrigeration cycle 
will operate in a dissipative mode, where power is dissipated to both the hot and cold baths.
\mbox{For an engine}, this dissipative mode will occur when the internal temperature of the oscillator 
Equation~(\ref{eq:inittemp})  after the expansion {\em adiabat} (cf. Figure \ref{fig:1} point D)
will exceed $T_h$ and  in a refrigerator cycle when the internal temperature exceeds $T_c$ 
(cf. Figure \ref{fig:3} point C). 

\subsection{Limit Cycle}
\label{sec:limit}

When a cycle is initiated, after a short transient time it settles to a steady-state operation mode. 
This periodic state is termed  the {\em limit cycle} \cite{k201,k116}.
An engine cycle converges to a limit cycle when the internal variables of the working medium reach a periodic steady state.
As a result, no energy or entropy is accumulated in the working medium. Figure \ref{fig:HLC} is an example of a periodic limit cycle.
Subsequently, a balance is obtained between the external driving and dissipation.
When the cycle time is reduced, friction causes additional heat to be accumulated in the working medium.
The cycle adjusts by  increasing the temperature gap between the working medium and the baths, leading to increased dissipation.
Overdriving leads to a situation where heat is dissipated to both the hot and cold bath and  power is only consumed.
When this mechanism is not sufficient to stabilise the cycle, one can expect a~breakdown of the concept of a limit cycle, resulting
in catastrophic consequences \cite{insinga2016thermodynamical}.

The properties of a completely positive (CP) map can be used to prove the existence of a limit cycle.
Lindblad \cite{lindblad75} has proven that the conditional entropy decreases when applying a trace-preserving
completely positive map  $\Lambda$   to both the state $\hat \rho $ and the reference state $\hat \rho _{ref}$:
$$ D( \Lambda \hat \rho || \Lambda \hat \rho_{ref} ) \le D(  \hat \rho ||  \hat \rho_{ref} )~,$$
where $D(\hat \rho || \hat \rho' )= Tr(\hat \rho(\log \hat \rho - \log \hat \rho'))$ is the conditional entropy distance.
A CP  map reduces the distinguishability between two states. This can be employed to prove the monotonic approach to steady-state,
provided that the reference state $\hat \rho_{ref}$ is the only invariant of
the CP map $\Lambda$ (i.e., \mbox{$\Lambda \hat \rho_{ref}= \hat \rho_{ref}$) \cite{frigerio1977quantum,frigerio1978stationary,spohn1978entropy}}.
This reasoning can prove the monotonic approach to the limit cycle. The~mapping imposed by the cycle of operation
of a heat engine is a product of the individual evolution
steps along the segments composing the cycle propagator.
Each one of these evolution steps is a completely
positive map, so the total evolution ${\cal U}_{cyc}$  Equation (\ref{eq:cycprop}) that
represents one cycle of operation is also a CP
map. If then a state $\hat \rho_{lc}$   is found that is a single invariant
of ${\cal U}_{cyc}$ (i.e., ${\cal U}_{cyc} \hat \rho_{lc}= \hat \rho_{lc}$), then any initial state $\hat \rho_{init}$ will monotonically
approach the limit cycle.

The largest eigenvalue of ${\cal U}_{cyc}$ with a value of $\bf 1$ is associated
with the invariant limit cycle state ${\cal U}_{cycr} \hat \rho_{lc}=1 \hat \rho_{lc}$, the
fixed point of ${\cal U}_{cyc}$. The other eigenvalues determine the rate of approach to the limit cycle.

In all cases studied of a reciprocating quantum heat engine, a single  non-degenerate eigenvalue of~$\bf 1$ was the only case found.
The theorems on trace preserving completely positive maps are all based on $C^*$ algebra, which means that the dynamical algebra of the system is compact.
Can the results be generalized to discrete non-compact cases such as the harmonic oscillator? In his study of the Brownian harmonic oscillator, Lindblad 
conjectured: ``{\em in the present case of a harmonic oscillator, the condition that ${\cal L}$ is bounded cannot
hold. We will assume this form for the generator with $\Op H$ and $\cal L$ unbounded as the simplest
way to construct an appropriate model}''
~\cite{lindblad1976brownian}.  
The master equation in Lindblad's form Equation (\ref{eq:dissipative}) is well established. Nevertheless, the non-compact character of the
resulting map has not been challenged. 

A  nice demonstration is the study of Insinga et al. \cite{insinga2016thermodynamical}, which shows conditions where a limit cycle is
not obtained. This study contains an extensive investigation of the limit cycles as a function of the parameters
of the system \cite{insinga2016thermodynamical}.

\subsection{Engine Operation and Performance}

The engine's cycle 
can operate in different modes, which are: 
adiabatic,  frictionless, friction-dominated,  and the sudden cycle.
In addition, one has to differentiate between two limits: high~temperature $k_B T \gg \hbar \omega $, 
where the unit of energy is $k_BT$, to low temperature $ \hbar \omega \gg k_B T$, where the unit of energy is $\hbar \omega$.

In the adiabatic and frictionless cycles \cite{del2014more}, the performance can be completely 
determined by the value of  energy at the switching point between strokes.

\subsubsection{Optimizing the Work per Cycle}

The adiabatic limit with infinite time allocations on all segments maximises the work. 
No~coherence is generated, and therefore the cycle can be described by the change in energy.
On the expansion {\em adiabat} $E_B =\frac{\omega_c}{\omega_h} E_A$, and on the compression {\em adiabat} $E_D =\frac{\omega_h}{\omega_c} E_C$.
As a result, when the cycle is closed, the heat transferred to the hot bath ${\cal Q}_h=E^A-E^D$ 
and to the cold bath ${\cal Q}_c=E_B-E_C$ are related: $\frac{{\cal Q}_c}{{\cal Q}_h}= \frac{\omega_c}{\omega_h}$.

The efficiency for an engine becomes the Otto efficiency:
\begin{equation}
\eta= \frac{\cal W}{{\cal Q}_h} = 1 -\frac{\omega_c}{\omega_h} \le 1- \frac{T_c}{T_h}~.
\label{eq:effic}
\end{equation} 

Choosing the compression ratio ${\cal C}= \frac{\omega_h}{\omega_c}=\frac{T_h}{T_c}$ maximises the work
and leads to Carnot efficiency  $\eta_o=\eta_c$.
Since for this limit the cycle time $\tau_{cyc}$ is infinite, the power ${\cal P} ={\cal W}/\tau_{cyc}$ of this cycle is obviously  zero.

\subsubsection{Optimizing the Performance of the Engine for Frictionless Conditions}

Frictionless solutions allow finite time cycles with the same efficiency $\eta_o= 1 - \frac{\omega_c}{\omega_h}$ as
the adiabatic case. 
A different viewpoint is to account as wasted work the average energy invested
in achieving the frictionless solution, termed {\em superadiabatic drive} 
 \cite{abah2016performance}:
\begin{equation}
\eta = \frac{\cal W}{ {\cal Q}_h + \langle H_{ch}\rangle +\langle H_{hc}\rangle }~,
\label{eq:abe}
\end{equation}
where $\langle H_{ch}\rangle$ is the average additional energy during the {\em adiabatc} stroke.
Using for example Equation~(\ref{eq:adprop}), the average additional energy becomes
$\langle H_{ch}\rangle = \frac{\omega_h}{\omega_c}E_C\frac{\mu^2}{4-\mu^2}$, which vanishes
as the non-adiabatic parameter $\mu \rightarrow 0$.
This additional energy $\langle H_{ch}\rangle$ in the engine is the price for generating coherence. Coherence
is exploited to cancel friction. This extra energy is not dissipated, and can therefore be viewed as a catalyst.
For this reason, we do not accept the viewpoint  of  \cite{abah2016performance}. 

In our opinion, what should be added to the accounting is the additional energy generated by noise on the controls:
\begin{equation}
\eta = \frac{\cal W}{ {\cal Q}_h + E_C \delta_{ch} +E_A \delta_{hc} } \le \eta_o~,
\label{eq:abe1}
\end{equation}
where $\delta_{hc}= \delta_a+\delta_f$ for the power {\em adiabat} and $\delta_a$ and $\delta_f$ 
are generated by amplitude and phase noise on the controller (cf. Section \ref{subsec:noise}).
A similar relation is found for the compression {\em adiabat}.

Optimizing power requires a finite cycle time $\tau_{cyc}$. 
Optimisation is carried out with respect to the time allocations
on each of the engine's segments:
$\tau_h$, $\tau_{hc}$, $\tau_c$, and $\tau_{ch}$. This sets the total cycle time
$\tau_{cyc} = \tau_h+\tau_{hc}+\tau_c+\tau_{ch}$.
The time allocated to the {\em adiabats} is constrained by the frictionless solutions $\tau_{hc}^*$ and $\tau_{ch}^*$. The resulting optimization is very close to the unconstrained optimum \cite{insinga2016thermodynamical}, especially in the interesting limit of low temperatures.
The frictionless conditions are obtained either from Equation~(\ref{eq:mucrit}) 
or from other shortcuts to adiabaticity  methods Equation  (\ref{eq:invar}). 
In the frictionless regime, the number operator is fixed at both ends of the {\em adiabat}. 
The main task is therefore to optimize the time allocated to thermalisation on the {\em isochores}. 
This heat transport is the source of entropy production.

The time allocations on the
{\em isochores} determine the change in the number operator 
$N=\langle\Op N \rangle =\frac{1}{\hbar \omega}\langle \Op H \rangle$ (cf. Equation (\ref{eq:motion})):
$N^{\bf B} = e^{-\Gamma_h \tau_h} \left( N^{\bf A} - N_{eq}^h \right) +N_{eq}^h $ 
on the hot {\em isochore}, 
where $N^{\bf B}$ is the number expectation value at the end of the hot {\em isochore}, 
$N^{\bf A}$ at the beginning, and $N_{eq}^h$ is the equilibrium value point {\bf E}. 
A similar expression exists for the cold {\em isochore}. 

\begin{figure}
\center{\includegraphics[height=5cm]{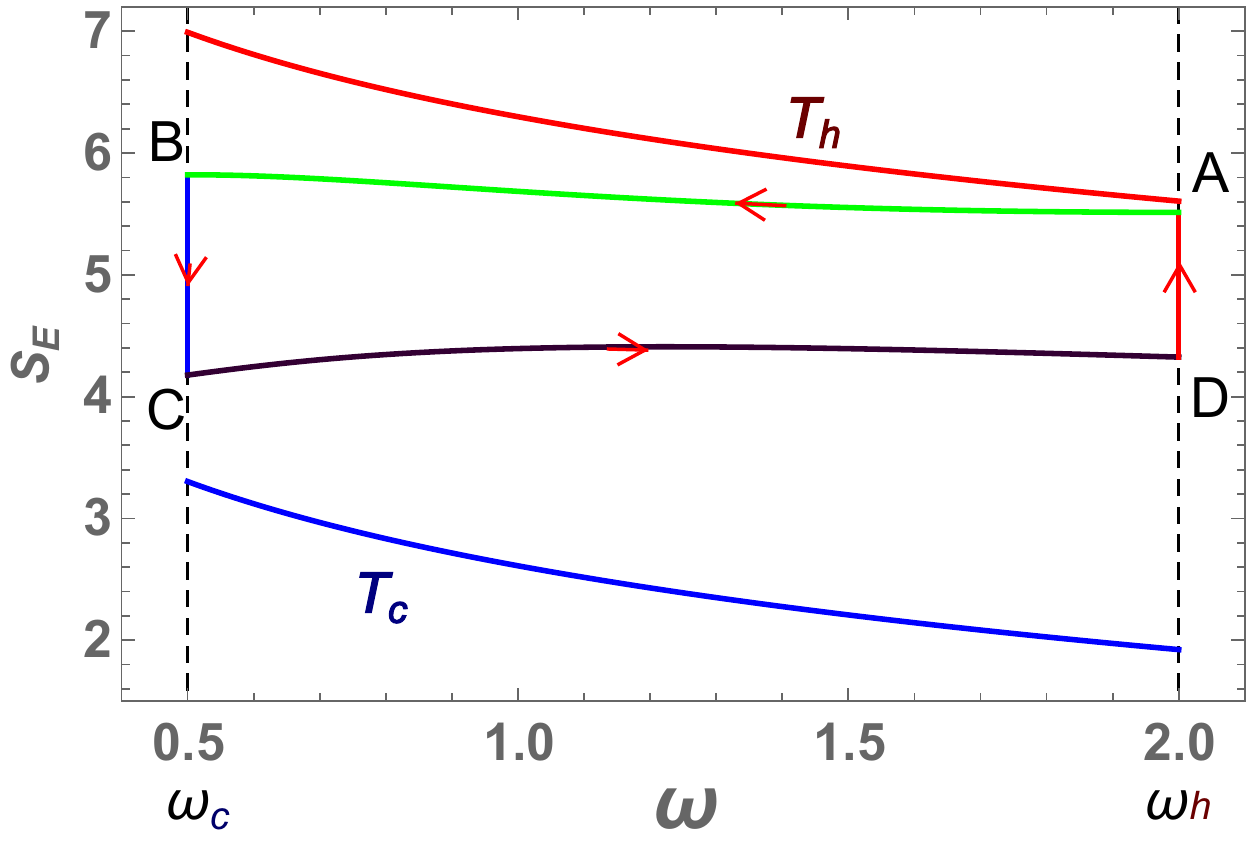}} 
\caption{{Typical engine cycle} ${\cal S}_E~$ vs. $~\omega$. 
Expansion {\em adiabat} $A \rightarrow B$. Cold {\em isochore} $B \rightarrow C$.
Compression {\em adiabat} $C \rightarrow D$. Hot {\em isochore} $ D \rightarrow A$.
The hot and cold isotherms are indicated.
The cycle parameters are  $\omega_c=0.5, ~T_c=5, ~\omega_h=2, ~T_h=200,~ \tau_c=\tau_h=2.1, ~|\mu|=0.8 ~\Gamma=1$.
} 
\label{fig:1}   
\end{figure}

Work in the limit cycle becomes
\begin{eqnarray}
{\cal W}_q ~~= ~~ E^{\bf C}-E^{\bf B}+E^{\bf A}-E^{\bf D}
~~  = ~~ \hbar(\omega_{c}-\omega_{h})(N^{\bf B}-N^{\bf D})~,
\label{eq:11}
\end{eqnarray}
where the convention of the sign of the work for a working engine is negative, in correspondance with Callen \cite{callen}, 
and we use the convention of Figure \ref{fig:1} to mark the population and energy at the corners of the cycle.

The heat transport from the hot bath becomes
\begin{eqnarray}
{\cal Q}_h ~~= ~~ E^{\bf B}-E^{\bf D}
~~  = ~~ \hbar\omega_{h}(N^{\bf B}-N^{\bf D})~.
\label{eq:heat11}
\end{eqnarray}

In the limit cycle for frictionless conditions, $N^{\bf B}=N^{\bf A}$, which leads to the relation
\begin{equation}
N^{\bf B}=\frac{(e^{\Gamma_{c}\tau_{c}}-1)}{1- e^{\Gamma_{c}\tau_{c}
+\Gamma_{h}\tau_{h}}}(N_{eq}^{h}-N_{eq}^{c})+N_{eq}^{h}~.
\label{eq:number1}
\end{equation}

In the periodic limit cycle,  the number operator change $N^{\bf B}-N^{\bf D}$ is equal on the hot and cold {\em isochores},
leading to the work per cycle:
\begin{eqnarray}
\label{eq:quasistatic work}
{\cal W}_q & = & \hbar(\omega_{h}-\omega_{c})(N_{eq}^{h}-N_{eq}^{c})\frac{(e^{x_{c}}-1)(e^{x_{h}}-1)}
{1-e^{x_{c}+x_{h}}}\label{eq:quasistatic-work}\\
 & \equiv & -G_W(T_{c},\omega_{c},T_{h},\omega_{h})F(x_{c},x_{h})~,
\nonumber
\end{eqnarray}
where the scaled time allocations are defined $x_{c}\equiv\Gamma_{c}\tau_{c}$ and $x_{h}\equiv\Gamma_{h}\tau_{h}$.
The work ${\cal W}_q$ Equation  (\ref{eq:quasistatic-work}) becomes a product 
of two functions: $G_W$, which is a function of the static constraints of the engine,
and $F$, which describes the heat transport on the {\em isochores}. Explicitly, the function $G_W$ is
\begin{equation}
G_W(T_{c},\omega_{c},T_{h},\omega_{h})=\frac{\hbar}{2}\left(\omega_{h}-\omega_{c}\right )
\left( \coth\left(\frac{\hbar\omega_h}{2 k_B T_h}\right)-\coth\left(\frac{\hbar\omega_c}{2 k_B T_c}\right)\right)~.
\label{eq:gfunct}
\end{equation} 

The function $F$ in Equation (\ref{eq:quasistatic work}) is bounded $ 0 \le F \le 1$; therefore,
for the engine to produce work, $G_W \ge 0$. 
The first term in (\ref{eq:gfunct}) is positive. Therefore, $G_W \ge 0$ requires 
that  $\frac{\omega_c}{\omega_h} \ge \frac{T_c}{T_h}$, 
or in terms of the compression ratio, $ 1 \le {\cal C} \le \frac{T_h}{T_c}$.
This is equivalent to the  statement that the maximum efficiency of the Otto cycle 
is smaller than the Carnot efficiency $\eta_{o} \le \eta_{c}$. 

In the high temperature limit when
$\frac{ \hbar \omega}{ k_BT} \ll 1$, $G_W$ simplifies to
\begin{equation}
G_W ~~=~~ k_B T_c(1-{\cal C})~+~k_BT_h(1-{\cal C}^{-1})~. 
\label{eq:hightempG}
\end{equation}

In this case, the work ${\cal W}_q = -G_W ~F$ can be optimized 
with respect to the compression ratio ${\cal C} = \frac{\omega_h }{\omega_c}$  
for  fixed bath temperatures. The optimum is found at ${\cal C}=\sqrt{\frac{T_h}{T_c}}$. 
As a  result, the efficiency at maximum power for high temperatures becomes
\begin{equation} 
\eta_q = 1-\sqrt{\frac{T_c}{T_h}}~,
\label{eq:endoeffic}
\end{equation}
which is  the well-known efficiency at maximum power of an endo-reversible engine 
\cite{novikov1958efficiency,curzon75,esposito2009universality,esposito2010efficiency,k85,k24}. 
Note that these results indicate greater validity 
to the Novikov--Curzon--Ahlbourn
~result from what their original derivation \cite{curzon75} indicates.

The function $F$ defined in (\ref{eq:quasistatic work}) characterizes the heat transport to the working
medium. As~expected, $F$ maximizes when infinite time is allocated to the {\em isochores}.
The optimal partitioning of the time allocation between the hot and cold {\em isochores}
is obtained when:
\begin{equation}
\Gamma_{h}(\cosh(\Gamma_{c}\tau_{c}) -1 )=\Gamma_{c}(\cosh(\Gamma_{h}\tau_{h})-1)~.
\label{eq: d-tau optimum work}
\end{equation}

If (and only if) $\Gamma_{h}=\Gamma_{c}$,
the optimal time allocations on the {\em isochores} becomes  $\tau_{h}=\tau_{c}$. 

Optimising the total cycle power output ${\cal P}$ is equivalent to optimizing $F/\tau_{cyc}$, since
$G_W$ is determined by the engine's external constraints. 
The total time allocation $\tau_{cyc}=\tau_{iso}+\tau_{adi}$ is partitioned to the time on the
{\em adiabats} $\tau_{adi}$, which is limited by the adiabatic frictionless condition,
and the time $\tau_{iso}$ allocated to the {\em isochores}. 

Optimising the time allocation on the {\em isochores} subject to (\ref{eq: d-tau optimum work}) leads to the optimal condition
\begin{equation}
\Gamma_{c}\tau_{cyc}(\cosh(\Gamma_{h}\tau_{h})-1)=
\sinh(\Gamma_{h}\tau_{h}+\Gamma_{c}\tau_{c})-\sinh(\Gamma_{c}\tau_{c})-\sinh(\Gamma_{h}\tau_{h})~.
\label{eq:quasistatic optimal tau}
\end{equation}

When $\Gamma_{h}=\Gamma_{c}\equiv\Gamma$, this expression simplifies to:
\begin{equation}
2x+\Gamma\tau_{adi}=2\sinh(x)~,
\label{eq:ftf}
\end{equation}
where $x=\Gamma_{c}\tau_{c}=\Gamma_{h}\tau_{h}$. For small $x$, Equation (\ref{eq:ftf})
can be solved, leading to the optimal time allocation on the {\em isochores}: 
$\tau_c=\tau_h \approx \left(\Gamma \tau_{adi}/3\right)^{\frac{1}{3}}/\Gamma$. Considering the restriction due to frictionless condition \cite{k242},  this time can be estimated to be:
$\tau_c=\tau_h \approx \frac{1}{\Gamma} \left(\frac{\Gamma}{\sqrt{\omega_c \omega_h}} \right)^{\frac{1}{3}}$.
When the heat transport rate $\Gamma$ is sufficiently large, the optimal power
conditions lead to the bang-bang solution where vanishingly small time is allocated
to all segments of the engine \cite{k116} and $\tau_{cyc} \approx 2 \tau_{adi}$.

The entropy production $\Delta S_U$  reflects the irreversible character of the engine.
In frictionless conditions, the irreversibility is completely associated with the
heat transport. $\Delta S_U$ can also be factorized to a product of two functions:
\begin{equation}
\Delta S_u ~~=~~G_S (T_{c},\omega_{c},T_{h},\omega_{h}) F(x_{c},x_{h})~,
\label{eq:entroporod}
\end{equation}
where $F$ is identical to the $F$ function defined in (\ref{eq:quasistatic-work}).
The function $G_S$ becomes:
\begin{equation}
G_S (T_{c},\omega_{c},T_{h},\omega_{h}) ~~=~~\frac{1}{2}
\left(\frac{\hbar \omega_h}{k_B T_h}-\frac{\hbar \omega_c}{k_B T_c} \right)
\left(\coth\left(\frac{\hbar\omega_c}{2 k_B T_c}\right) -\coth\left(\frac{\hbar\omega_h}{2 k_B T_h}\right)\right)~.
\label{eq:entroporod2}
\end{equation}

Due to the common $F(x_{c},x_{h})$ function, the entropy production has the same dependence 
on the time allocations $\tau_h$ and $\tau_c$ as the work ${\cal W}$ \cite{k152}.
As a consequence, maximizing the power will also maximize the entropy production 
rate $\Delta S_u/\tau_{cyc}$. Note that entropy production is {\em always} positive, even~for cycles that produce no work, as their compression ratio $\cal C$ is too large, which is a statement of the second law of thermodynamics.

The dependence of the $G_s$ function on the compression ratio can be simplified
in the high temperature limit, leading to:
\begin{equation}
G_S ~~=~~ {\cal C} \frac{T_c}{T_h}+{\cal C}^{-1}\frac{T_h}{T_c}-2~,
\label{eq:entroporod3}
\end{equation}
which is a monotonic decreasing function in the range $ 1 \le {\cal C} \le \frac{T_h}{T_c}$ 
that reaches a minimum at the Carnot boundary when ${\cal C}=\frac{T_h}{T_c}$.
When power is generated, the entropy production rate in the frictionless engine is linearly proportional to the power:
\begin{equation}
{\cal S}_u = \left(\frac{\hbar \omega_h}{k_B T_h}-\frac{\hbar \omega_c}{k_B T_c} \right)\left(\frac{1}{\omega_h-\omega_c}\right){\cal P}~.
\label{eq:entpow}
\end{equation}

Frictionless harmonic cycles have been studied under the name of {\em superadiabatic driving} \cite{del2014more}.
The~frictionless {\em adiabats} are obtained using the methods of {\em shortcut to adiabaticity} \cite{muga2016}
and the invariant Equation (\ref{eq:invar}). An important extension applies
shortcuts to adiabaticity  to working mediums composed of interacting particles in a harmonic trap 
\cite{wang2015efficiency,jaramillo2016quantum,beau2016scaling,chotorlishvili2016superadiabatic}. 

A variant of the Otto engine is an addition of projective energy measurements before and after each {\em adiabat}. 
This construction is added to measure the work output \cite{zheng2016}.
As a result, the working medium is always diagonal in the energy basis.
In the frictionless case, the cycle is not altered by this projective measurement of energy.

\subsubsection{The Engine in the Sudden Limit}

The extreme case of the performance of an engine with zero time allocation on the {\em adiabats} 
is dominated by the {\em frictional} terms. These terms arise from the inability
of the working medium to adiabatically  follow the external change in potential. 
A closed-form expression for the sudden limit can be derived based on the 
{\em adiabatic} branch propagator ${\cal U}_{hc}$ and ${\cal U}_{ch}$ in
Equation (\ref{eq:sudden}). 

To understand the role of friction, we demand that the heat conductance terms $\Gamma$ are very large,
thus eliminating the thermalisation time.
In this limiting case, the work per cycle becomes:
\begin{equation}
{\cal W}_s ~~=~~ \frac{(\omega_c -\omega_h)(\omega_c +\omega_h)}{4 \omega_c\omega_h}
 \left(\hbar \omega_c \coth(\frac{\hbar \omega_h}{2 k_B T_h})
-\hbar \omega_h \coth(\frac{\hbar \omega_c}{2 k_B T_c})\right)~.
\label{eq:sudwork1}
\end{equation}

The maximum produced work $-{\cal W}_s$ can be optimised with respect to the compression ratio $\cal C$.
At~the high temperature limit: 
\begin{equation}
{\cal W}_s = \frac{1}{2}k_B T_h ( {\cal C}^2-1)(\frac{T_c}{T_h}-\frac{1}{{\cal C}^2})~.
\label{eq:sudworkh}
\end{equation}

For the frictionless optimal compression ratio ${\cal C}= \sqrt{ \frac{T_h}{T_c}}$, ${\cal W}_s$ is zero.
The optimal compression ratio for the sudden limit becomes: 
${\cal C}= \left( \frac{T_h}{T_c} \right)^{1/4}$, leading to
the maximal work in the high temperature~limit
\begin{equation}
{\cal W}_s ~~=~~ -\frac{1}{2} k_B T_c \left(~ 1-\sqrt{\frac{T_h}{T_c}}~ \right)^2~.
\label{eq:sudwork}
\end{equation}

The efficiency at  the maximal work point becomes:
\begin{equation}
\eta_s~~=~~ \frac{1 -\sqrt{\frac{Th}{Tc}}}{2+\sqrt{\frac{Th}{Tc}}}~.
\label{eq:sudefic}
\end{equation}

Equation (\ref{eq:sudefic}) leads to the following hierarchy of the engine's maximum work
efficiencies:
\begin{equation}
\eta_s ~\le ~\eta_q ~\leq~ \eta_c~.
\label{eq:efic}
\end{equation}

Equation (\ref{eq:efic}) leads to the interpretation that when  the engine is constrained by friction its efficiency
is smaller than the endo-reversible efficiency, where the engine is constrained by heat transport that is smaller than 
the ideal Carnot efficiency. At the limit of $T_c \rightarrow 0$, we have
$\eta_s = \frac{1}{2}$ and $\eta_q = \eta_c=1$ \cite{k292}.

An upper limit to the work invested in friction ${\cal W}_f$ is obtained by subtracting the 
maximum work in the frictionless limit Equation (\ref{eq:quasistatic work}) from the maximum work in the 
sudden limit Equation (\ref{eq:sudwork1}).
In~both these cases, infinite heat conductance is assumed, leading to $N^B=N_{eq}^h$ and $N^D=N_{eq}^c$. Then, 
the upper limit of work invested to counter friction becomes: 
\begin{equation}
{\cal W}_f ~~=~~ \hbar \omega_h \frac{({\cal C}-1)^2 (1+{\cal C}+2{\cal C} N_{eq}^c+2N_{eq}^h)}{4{\cal C}^2}
~.
\label{eq:wfric3}
\end{equation}

At high temperature, Equation (\ref{eq:wfric3}) changes to:
\begin{equation}
{\cal W}_f ~~=~~\frac{1}{2}k_b T_h ({\cal C}-1)^2 ({\cal C}^{-2} + \frac{T_c}{T_h})
~. 
\label{eq:wfrichigh}
\end{equation}

The maximum produced work at the high temperature limit of
the frictionless and sudden limits differ by the optimal compression ratio. For the
frictionless case, ${\cal C}^*= \sqrt{\frac{T_h}{T_c}}$, and 
for the sudden case, ${\cal C}^*= (\frac{T_h}{T_c})^{1/4}$.

The work against friction ${\cal W}_f$ {(}Equation (\ref{eq:wfrichigh})) is an increasing function of the temperature ratio.
For~the compression ratio that optimises the frictionless limit, the sudden work is zero.
At this compression ratio, all the useful work is balanced by the work against friction ${\cal W}_f={\cal W}_q$.
Beyond this limit, the engine transforms to a dissipator, generating entropy at both the hot
and cold baths.
This is in contrast to the frictionless limit, where the compression ratio ${\cal C}=\frac{T_h}{T_c}$ 
leads to zero power. 

The complete sudden limit assumes short time dynamics on  all segments including the {\em isochores}. 
These cycles with vanishing cycle times approach the  limit of a continuous engine.
The short time on the {\em isochores} means that coherence can survive. 
Friction can be partially avoided by exploiting this coherence, which---unlike the frictionless engine---is present in the four corners of the cycle.
The condition for such cycles is that the time allocated 
is much smaller than the natural period set by the frequency $\tau_c, \tau_h \ll 2 \pi/\omega$ 
and by heat transfer  $\tau_c, \tau_h \ll 1/\Gamma$. 
The heat transport from the hot and cold baths in each stroke becomes very small.
For simplicity, $\Gamma_h \tau_h=\Gamma_c \tau_c$ is 
~chosen to be balanced. 
Under these conditions, the cycle propagator becomes:
\begin{eqnarray}
{\cal U}_{cyc}~=~
\left(
\begin{array}{cccc}
(1-g)^2&0&0&[g(1-g)\frac{1}{2}(1+{\cal C}^2)\hbar \omega_c N_c^{eq} +g \hbar \omega_h N_h^{eq}]\\
0&(1-g)^2&0&(1-g)g\frac{1}{2}(1-{\cal C}^2)\hbar \omega_c N_c^{eq} \\
0&0&(1-g)^2&0\\
0&0&0&1
\end{array}
\right)~,
\label{eq:sprop}
\end{eqnarray}
where  the degree of thermalisation is $g=1-R \approx \Gamma_h \tau_h$.
Observing Equation (\ref{eq:sprop}), it is clear that the limit cycle vector contains both $\Op H$ and $\Op L$.

The work output per cycle becomes:
\begin{equation}
{\cal W}_S=- \hbar \omega _h \frac{g}{2-g}\frac{{\cal C}^2-1}{2 {\cal C}^2}\left( N_h^{eq}-  {\cal C} N_c^{eq}\right)~.
\end{equation}

Extractable work is obtained in the compression range of $1 < {\cal C} < \frac{N_h^{eq}}{N_c^{eq}}$.
The maximum work is obtained when ${\cal C}^* = \left( \frac{N_h^{eq}}{N_c^{eq}} \right)^{1/4}$.
At high temperature, the work per cycle simplifies to:
\begin{equation}
{\cal W}_S \approx -k_B T_c \frac{\Gamma \tau } {2 } \frac{{\cal C}^2-1}{2 {\cal C}^2}\left( \frac{T_h}{T_c}-{\cal C}^2 \right)~.
\end{equation}

The work vanishes for the frictionless compression ratio ${\cal C}= \sqrt{ \frac{T_h}{T_c}}$.
The optimal compression ratio at the high temperature limit becomes: 
${\cal C}^* =\left( \frac{T_h}{T_c}\right)^{1/4}$.

The entropy production becomes:
\begin{equation}
\Delta S_u = \hbar \omega_h \frac{g}{2-g}\frac{{\cal C}^2-1}{2 {\cal C}}\left( N_h^{eq}(\frac{1+{\cal C}^2}{2 {\cal C} T_c} -\frac{\cal C}{T_h})+  N_c^{eq}(\frac{1+{\cal C}^2}{2 T_h} -\frac{1}{T_c})\right)~.
\end{equation}

Even for zero power (e.g., ${\cal C}=1$), the entropy production is positive, reflecting a heat leak
from the hot to cold bath.

The power of the engine for zero cycle time $\tau_h \rightarrow 0$ and $\tau_c \rightarrow 0$ is finite:
\begin{equation}
{\cal P}_S = -\frac{\hbar \omega_h \Gamma }{2} \frac{{\cal C}^2-1}{2 {\cal C}^2}\left( N_h^{eq} -  {\cal C} N_c^{eq}\right)~.
\end{equation}

This means that we have reached the limit of a continuously operating engine.
This  observation is in accordance with the universal limit of small action on each segment \cite{k299,k306}.
When additional dephasing is added to Equation (\ref{eq:sprop}), no useful power is produced and the cycle operates in
a~dissipator~mode.

The efficiency of the complete sudden engine becomes:
\begin{equation}
\eta_S   
=\frac{{\cal C}^2-1}{2 {\cal C}^2}\frac{1-{\cal C}\frac{N_c^{eq}}{N_h^{eq}}}{2-(1+{\cal C}^2)\frac{1}{\cal C}\frac{N_c^{eq}}{N_h^{eq}}}~.
\end{equation}

The extreme sudden cycle is a prototype of a quantum phenomenon
an engine that requires global coherence to operate.
At any point in the cycle, the working medium state is non-diagonal in the energy~representation.

\subsubsection{Work Fluctuation in  the Engine Cycle}

Fluctuations are extremely important for a single realisation of a quantum harmonic engine. 
The~work fluctuation can be calculated from the fluctuation of the energy 
at the four corners of the cycle \cite{funo2016universal,seifert2012stochastic}. The energy fluctuations for a generalised Gibbs state {(}Equation (\ref{eq:product}))
is related to the internal temperature {(}Equation (\ref{eq:inittemp})) $Var (E) =(k_B T_{int})^2$.
For frictionless cycles, the variance of the work~becomes:
\begin{equation}
Var ({\cal W}) = (k_B T_{int}^h)^2(1+\frac{1}{{\cal C}^2}) + (k_B T_{int}^c)^2(1+{\cal C}^2)~,
\label{varwork}
\end{equation}
where $T_{int}^h$ and $T_{int}^c$ are the internal temperatures at the end of the hot and cold  thermalisation.
For~the case of complete thermalisation when the oscillator reaches the temperature of the bath, the work variance
is smallest for the Carnot compression ratio ${\cal C} = T_h/T_c$.
Generating coherence will increase the energy variance (cf. Equation (\ref{eq:var})), and with it the work variance \cite{funo2016universal}.

\subsubsection{Quantum Fuels: Squeezed  Thermal Bath}

Quantum fuels represent a resource reservoir that is not in thermal equilibrium due to quantum coherence or quantum correlations.
The issue is how to exploit the additional out-of-equilibrium properties of the bath.
The basic  idea of quantum fuels comes from the understanding that coherence can reduce the von Neumann entropy of the fuel. 
In principle, this entropy can  be exploited to increase the efficiency of the engine without violating the second-law \cite{scully2003extracting}.
An example of such a fuel is supplied by a squeezed thermal bath
\cite{rossnagel2014nanoscale,abah2014efficiency,galve2009nonequilibrium,manzano2016entropy,manzano2015perfect,li2016non,niedenzu2015efficiency,niedenzu2017universal}.
Such a bath delivers a combination of heat and coherence. As a result,  work can be extracted 
from a single heat bath without violating the laws of thermodynamics.
An additional suggestion for a quantum fuel is a non-Markovian hot bath \cite{zhang2014quantum}.

The model of this engine starts from
a squeezed boson hot bath where $\Op H_B =\sum_k \hbar \Omega_k \Op b^{\dagger}_k \Op b_k$. 
This~bath is coupled  to the working medium by the interaction 
$\Op H_{SB}= \sum_k i g_k (\Op a \Op b^{\dagger}_k - \Op a^{\dagger} \Op b_k)$. 
As a result, the master equation describing thermalisation  {(}Equation (\ref{eq:dissipative})) is modified to \cite{manzano2016entropy,li2016mutual}:
\begin{equation}
\mathcal{L}_{D}(\hat \rho)
~~=~~k_{\uparrow}(\Op s^{\dagger}\hat \rho \Op {s}~-~\frac{1}{2}\{\Op {s}\Op {s}^{\dagger},\hat \rho\})+
k_{\downarrow}(\Op{s}\hat \rho \Op {s}^{\dagger}-\frac{1}{2}\{\Op{s}^{\dagger}\Op{s},\hat \rho\})~,
\label{eq:sqbath}
\end{equation}
where $\Op s =\Op a \cosh(\gamma) + \Op a^{\dagger} \sinh(\gamma) = \Op S \Op a \Op S^{\dagger}$. 
$\Op S$ is the squeezing operator {(}Equation (\ref{eq:sqeez})) and $\gamma$ the squeezing parameter.

Under squeezing,  the equation of motion of the hot {\em isochore} thermalisation {(}Equation (\ref{eq:motion})) is 
~modified to:
\begin{eqnarray}
\frac{d}{dt}\left(\begin{array}{c}
\Op H\\
\Op L\\
\Op C\\
\Op I
\end{array}\right)(t)=
\left(\begin{array}{cccc}
-\Gamma & 0 & 0 & \Gamma \langle \Op H \rangle_{sq}\\
0 & -\Gamma & -2\omega & 0 \\
0 & 2 \omega & -\Gamma & \Gamma \langle \Op C \rangle_{sq}\\
0&0&0&0
\end{array}\right)
\left(\begin{array}{c}
\Op H\\
\Op L\\
\Op C\\
\Op I
\end{array}\right)(t)~,
\label{eq:squeez1}
\end{eqnarray}
where $\Gamma=k_{\downarrow}-k_{\uparrow}$ is the heat conductance and 
$k_{\uparrow}/k_{\downarrow}= e^{-\hbar\omega_h/k_B T_h}$ obeys detailed balance. 
The~difference from  the normal thermalisation dynamics Equation (\ref{eq:motion}) is in
the equilibrium values: 
$\langle \Op H \rangle_{sq} = \cosh^2(\gamma)\langle \Op H \rangle_{eq} + \sinh^2(\gamma) \hbar \omega_h \frac{k_{\downarrow}}{\Gamma} $,
where $\langle \Op H \rangle_{eq}$ is  the equlibrium value of the oscillator at temperature $T_h$.
In addition, the invariant state of Equation (\ref{eq:squeez1}) contains coherence:
$\langle \Op C \rangle_{sq} = - \sinh( 2 \gamma ) \frac{k_{\uparrow}+k_{\downarrow}}{\Gamma} $.
This coherence is accompanied by additional energy that is transferred to the system.
The squeezed bath delivers extra energy to the working fluid as if the hot bath has a higher temperature, 
since $\langle \Op H \rangle_{sq} \ge \langle \Op H \rangle_{eq}$.
This temperature can be  calculated from Equation (\ref{eq:inittemp}). The thermalization to the squeezed bath generates mutual correlation between the system and bath \cite{li2016mutual}.

The  coherence transferred to the system $\langle \Op C \rangle \le \langle \Op C \rangle_{sq}$ can be cashed upon to increase
the work of the cycle. This requires an {\em adiabatic} protocol which is similar to the frictionless case.
In the frictionless case, the protocol of $\omega(t)$ was chosen to cancel the coherence generated during the stroke
and to reach a state diagonal in energy. This protocol can be modified to exploit the initial coherence
and to reach a~state diagonal in energy but with lower energy, thus producing more work. The coherence thus serves as a source of quantum availability, allowing more work to be extracted from the system \cite{hoffmann2015finite,hoffmann2015quantum,niedenzu2017universal}.
For example, using the propagator on the {\em adiabat } Equation (\ref{eq:adprop}) based on $\mu=constant$,
the stroke period $\tau_{hc}$ can be increased from the frictionless value to add a rotation $\cos(\Omega \Theta (t)) =\frac{\mu^2}{4}$,
which will null the coherence and reduce the final energy. 
Other frictionless solutions could be modified to reach the same~effect.

\subsection{Closing the Cycle: The Performance of the Refrigerator}

\begin{figure}
\center{\includegraphics[height=5.5cm]{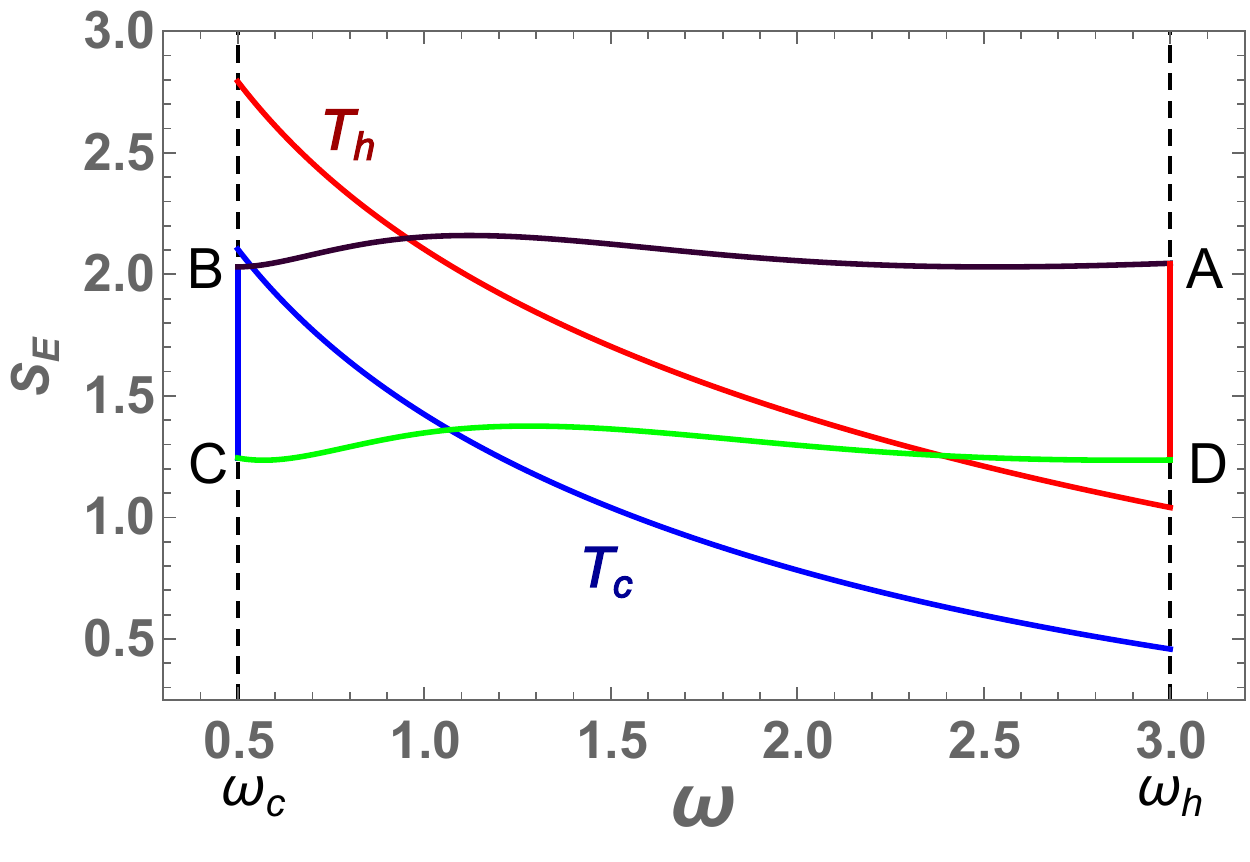}} 
\caption{ Typical frictionless refrigerator cycle ${\cal S}_E$ vs. $\omega$. 
Expansion {\em adiabat} $D \rightarrow C$. Cold {\em isochore} $C \rightarrow B$.
Compression {\em adiabat} $B \rightarrow A$. Hot {\em isochore} $ A \rightarrow D$.
The hot and cold isotherms are indicated.
The cycle parameters are $\omega_c=0.5, ~T_c=1.5, ~ \omega_h=3, ~T_h=3 , ~\tau_c=\tau_h=2.1, ~|\mu|=0.5 , \Gamma=1$.
} 
\label{fig:3}   
\end{figure}

A refrigerator or heat pump employs  the working medium to shuttle heat from
the cold to hot reservoir. A prerequisite for cooling is that the expansion {\em adiabat} should cause the
temperature of the working medium to be lower than the cold bath. In addition, at the end of the compression {\em adiabat},
the temperature should be hotter than the hot bath (cf. Figure \ref{fig:3}).
To generate a refrigerator, we use the order of stroke propagators in Equation (\ref{eq:cycref}). 
The heat extracted from the cold bath becomes:
\begin{equation}
{\cal Q}_c = E^{\bf B}-E^{\bf C} = \hbar \omega_c (N^{\bf B} - N^{\bf C})~.
\label{eq:heat}
\end{equation}

The interplay between efficiency and cooling power is the main theme in the performance analysis.
The efficiency of a refrigerator is defined by the coefficient of performance (COP):
\begin{equation}
COP = \frac{{\cal Q}_c}{\cal W} =\frac{\omega_c}{\omega_h-\omega_c} \le \frac{T_c}{T_h-T_c}~.
\label{eq:cop}
\end{equation}

The cooling power ${\cal R}_c$ is defined as:
\begin{equation}
{\cal R}_c =\frac{{\cal Q}_c}{\tau_{cyc}}~.
\label{eq:rc}
\end{equation}

Optimising the performance of the refrigerator can be carried out by a similar analysis to the one employed for the heat engine.
Insight into the ideal performance can be gained
~by examining the expansion {\em adiabat}.
The initial excitation should be minimized, requiring the hot bath to cool the working medium to its ground state.
This is possible if $\hbar \omega_h \gg k_B T_h$.
Next, the expansion should be as adiabatic as possible so that at the end the working medium is still 
as close as possible to its ground state $E_C \approx \frac{\hbar \omega_c}{2}$.
The frictionless solutions found in Section \ref{subsec:adidy} can be employed to achieve this task in minimum time.

\subsubsection{Frictionless Refrigerator}

The adiabatic refrigerator is obtained in the limit of infinite time $\mu \rightarrow 0$, leading to  
constant population $N$ and ${\cal S}_E$. Then, $E^{\bf C} = \frac{\omega_c}{\omega_h} E^{\bf D}$. 
At this limit, since $\tau \rightarrow \infty$, the cooling rate vanishes ${\cal R}_c=0$.
The Carnot efficiency can be obtained when ${\cal C} =\frac{T_h}{T_c}$.

Frictionless solutions require that the state $\hat \rho$ 
is diagonal in energy in the beginning and at the end of the {\em adiabat} .
The analytic propagator on the expansion {\em adiabat} {(}Equation (\ref{eq:adprop})) describes the expansion adiabat:
${\bf D} \rightarrow {\bf C}$ :
\begin{equation}
E^{\bf C} = \frac{ 1}{{\cal C}} ~\frac{1}{\Omega^2}\left(4- \mu^2 c \right)\cdot E^{\bf D}~,
\label{eq:hadi}
\end{equation}
where $c= \cos(\Omega \theta_c)$ and $\theta_c=-\frac{1}{\mu}\log\left({\cal C}\right)$.

Frictionless points are obtained whenever $N^{\bf C}=N^{\bf D}$.
The condition is $c=1$ in Equation (\ref{eq:hadi}). 
Then,~$\mu < 2$,  leading to the critical frictional points {(}Equation (\ref{eq:mucrit})).
These solutions have optimal efficiency Equation (\ref{eq:cop}) with finite power.
The optimal  time allocated to the {\em adiabat} becomes {(}cf. Equation (\ref{eq:tperiodic})) $\tau_{hc}^* = (1-{\cal C})/ (\mu^* \omega_h)$.

This frictionless solution with a minimum time allocation $\tau_{hc}^*$ 
scales as the inverse frequency $\omega_c^{-1}$, which outperforms the linear ramp solution $\omega(t)=\omega_i+gt$.

Other faster frictionless solutions can be obtained using the protocols of Section \ref{subsec:adidy}, such as
the superadiabatic protocol or by applying optimal control  theory \cite{k242}.
Both cases  lead to the scaling of the adiabatic expansion time as $\tau_{hc} \propto \frac{1}{\sqrt{\omega_c \omega_h}}$.

Once the time allocation on the {\em adiabats} is set, the
time allocation on the {\em isochores} is optimised for the thermalisation using the method of \cite{k221}, 
and the optimal cooling power becomes:
\begin{equation}
{\cal R}_c^* ~~=~~ \frac{e^z}{(1+e^z)^2}\Gamma \hbar \omega_c (N_c^{eq}-N_h^{eq})~,
\label{eq:rc-2}
\end{equation}
where $z=\Gamma_h \tau_h=\Gamma_c \tau_c$. The optimal $z$ is determined by the solution of the equation 
$2z+\Gamma (\tau_{hc}+\tau_{ch})=2 \sinh(z)$.

\subsubsection{The Sudden Refrigerator}

Short {\em adiabats} generally lead to the excitation of the oscillator and result in friction (cf. Section \ref{subsec:adidy}).
Nevertheless, a refrigerator can still operate at the limit of vanishing cycle time. 
In a similar fashion to the sudden engine, coherence can be exploited and leads to
a finite cooling power when $\tau_{cyc} \rightarrow 0$.
The~cooling power for the sudden limit becomes:
\begin{equation}
{\cal R}_c = -\hbar \omega_c \Gamma \left( N_c^{eq} -  \frac{1}{2 {\cal C}}(1+ {\cal C}^2)N_h^{eq}\right)~.
\end{equation}

Note that the cooling rate in this sudden-limit becomes zero at a sufficiently low cold bath temperature so that $N_c^{eq}=(1+{\cal C}^2) N_h^{eq}$. This formula loses its meaning and should not be used below this temperature.

\subsubsection{The Quest to Reach Absolute Zero}

The quantum harmonic refrigerator can serve as a primary model to explore cooling at very low temperatures. A necessary condition  is that the internal temperature of the oscillator
{(}Equation~(\ref{eq:inittemp})) should be lower than the cold bath. $T_{int} < T_c$ when $T_c \rightarrow 0$.
This condition imposes very high compression ratios ${\cal C} \gg \frac{T_h}{T_c}$  
so that at point $D$ (cf. Figure \ref{fig:3}), at the end of the hot thermalization, 
the~oscillator is very close to the ground state.

An important feature of the model is that the cooling power  vanishes as $T_c$ approaches zero.
Qualitatively, ${\cal R}_c \rightarrow 0$  means that the adiabatic expansion from point 
$D \rightarrow C$ for high compression ratios requires a significant amount of time. Another issue
is the rate of cold thermalisation $C \rightarrow B$ and its scaling with $T_c$ 
when the oscillator extracts heat from the cold bath.
These issues can be made quantitative
~by exploring the scaling exponent $\alpha$ of the optimal cooling power 
with the cold bath temperature $T_c$:
\begin{equation}
{\cal R}_c=\frac{{\cal Q}_c}{\tau_{cyc}} \propto T_c^{1+\alpha}~.
\label{eq:scatc}
\end{equation}

The vanishing of the cooling power ${\cal R}_c$ as $T_c \rightarrow 0$ is related to a dynamical version of the third-law of thermodynamics \cite{k275,k281}.

{Walther Nernst  formulated 
two independent formulations of the third-law of \mbox{thermodynamics~\cite{nerst06,nerst06b,nerst18}}. 
The first is a purely static (equilibrium) one, also known as the ``Nernst heat theorem'',
phrased}:
\begin{itemize}
\item{The entropy of any pure substance in thermodynamic equilibrium approaches zero as the temperature approaches
zero.}
\end{itemize}

\noindent {The second}
~formulation is  dynamical, known as the unattainability principle \cite{landsberg56,landsberg89,wheeler91,belgiorno03,k275}:
\begin{itemize}
\item{It is impossible by any procedure---no matter how idealised---to reduce any assembly to absolute zero temperature
in a finite number of operations \cite{nerst18}}.
\end{itemize}

The second law of thermodynamics already imposes a restriction on $\alpha$ \cite{k156,k275,k281}.
In steady-state,  the entropy production rate is positive. Since the process is cyclic, it takes place only in the baths:
$\sigma = \dot {\cal S}_c + \dot {\cal S}_h \ge 0$.
When the cold bath  approaches the absolute zero temperature,
it is necessary to eliminate the entropy production divergence at the cold side because 
$\dot {\cal S}_c =\frac{{\cal R}_c}{T_c} $. 
Therefore, the entropy production at the cold bath  when $T_c \rightarrow 0$  scales as: 
\begin{equation}
\dot {\cal S}_c \sim - T_c^{\alpha}~,~~~~\alpha \geq 0~.
\label{eq:III-1}
\end{equation}

For the case when $\alpha=0$, the fulfillment of the second-law depends on the entropy production of the other baths, 
which should compensate for the negative entropy production of the cold bath. 
The first formulation of the third-law slightly modifies this restriction. 
Instead of $\alpha \geq 0$, the third-law imposes $\alpha > 0 $, 
guaranteeing that  the entropy production at the cold bath is zero at  absolute zero: $\dot S_c = 0$. 
This requirement leads to the scaling condition of the heat current ${\cal R}_c \sim T_c^{\alpha+1}~,~~\alpha >0$.  

The second formulation of the third-law is a dynamical one, known as the unattainability principle:
 \emph{no refrigerator can cool a system to absolute zero temperature at finite time}.
This formulation is more restrictive, imposing limitations on the system bath interaction and the cold bath properties when $T_c \rightarrow 0$ \cite{k275}. 
The rate of temperature decrease of the cooling process should vanish according to the characteristic exponent $\zeta$: 
\begin{equation}
\label{eq:cp}
 \frac{dT_c(t)}{dt} \sim -T_c^{\zeta}, ~~~ T_c\rightarrow 0 ~~.
\end{equation}

In order to evaluate Equation (\ref{eq:cp}), the heat current can be related to the temperature change:
\begin{equation}
 {\cal J}_c(T_c(t)) = -c_V(T_c(t))\frac{dT_c(t)}{dt}~.
 \label{25}
\end{equation}

This formulation takes into account the heat capacity  $c_V(T_c)$ of the cold bath. $c_V(T_c)$ is determined by the 
behaviour of the degrees of freedom of the cold bath at low temperature.  Therefore, the scaling exponents
can be related $\zeta=1+\alpha - \eta$,
where $c_V \sim T_c^{\eta}$ when $T_c \rightarrow 0$.

The harmonic quantum refrigerator is a primary example to explore the emergence of quantum dynamical
restrictions that result in cooling power consistent with the third-law of thermodynamics.
Analysis of the {\em adibatic} expansion will lead to insight on the cooling rate ${\cal R}_c$ and the exponent $\alpha$. 

The frictionless solutions lead to an upper bound on the optimal cooling rate  {(}Equation (\ref{eq:rc})).
For~the limit $T_c \rightarrow 0$, $\Gamma \tau_{hc}$ is large; therefore, $z$ is large, leading to:
\begin{equation}
{\cal R}_c^* ~\approx~ \frac{\Gamma (\tau_{hc}+\tau_{ch})}{(1+\Gamma \tau_{hc})^2} \Gamma \hbar \omega_c (N_c^{eq}-N_h^{eq})~.
\label{eq:rc1}
\end{equation}

At high compression ratio, $N_h^{eq} \rightarrow 0$,  and in addition $\omega_c \ll \Gamma$ one obtains:
\begin{equation}
{\cal R}_c^* ~\approx~ \frac{1}{ \tau_{hc}} \Gamma \hbar \omega_c N_c^{eq}~.
\label{eq:rc2}
\end{equation}

Optimizing ${\cal R}_c$ with respect to $\omega_c$ leads to a linear relation between $\omega_c$ and $T_c$, 
$\hbar \omega_c = k_B T_c$; therefore:
\begin{equation}
{\cal R}_c \le A \omega^{\nu} N_c^{eq}~,
\end{equation}
where $A$ is a constant and the exponent $\nu$ is either $\nu=2$ for the $\mu=const$ solution or $ \nu=\frac{3}{2}$
for the optimal control solution. Therefore:
\begin{equation}
{\cal R}_c^* ~\approx~  \hbar \omega_c^2 N_c^{eq}
\label{eq:rate1}
\end{equation}
for the $\mu=const$ frictionless solution, and 
\begin{equation}
{\cal R}_c^* ~\approx~  \frac{1}{2}\hbar \omega_c^{\frac{3}{2}} \sqrt{\omega_h} N_c^{eq}
\label{eq:rate2}
\end{equation}
for the optimal control frictionless solution.
Due to the linear relation between $\omega_c$ and $T_c$, Equations~(\ref{eq:rate1}) and (\ref{eq:rate2}) determine the exponent $\alpha$, where $\alpha=1$ for the frictionless scheduling with constant $\mu$, 
and $\alpha=\frac{1}{2}$ for the optimal control frictionless scheduling.
In all cases, the dynamical version of Nernst's heat law is observed based only on the adiabatic expansion.

If one is forced to spend less time on the {\em adiabat} than the minimal time required for a shortcut solution, the oscillator cannot reach arbitrarily low energies or temperatures at the end of the expansion~\cite{stefanatos2010,stefanatos2016minimum}. 
At the limit, one approaches the sudden {\em adiabat} limit. In this case, the refrigerator cannot cool below a minimal ($T_c^*>0$) temperature, and the refrigerator thus satisfies the unattainability principle trivially.

The unattainability principle is related to the scaling of the heat transport $\Gamma_c$ with $T_c$. 
This issue has been explored in \cite{k275,k281}, and is related to the scaling of the heat conductivity with temperature.
The arguments of \cite{k275} are applicable to the quantum harmonic refrigerator.

\section{Overview}

Learning from example has been one  of the major sources of insight in the study of thermodynamics.
A good example can bridge the gap between concrete and abstract theory.
The harmonic oscillator quantum Otto cycle serves as a primary example of a quantum thermal device 
inspiring  experimental realisation \cite{Rossnagel325,blickle2012realization}. 
On the one hand, the model is very close to actual physical realisations  in many scenarios \cite{zhang2014prl}.
On the other hand, many features of the model can be obtained as closed-form analytic solutions.

Many of the features obtained for the quantum harmonic Otto engine have been observed in
stochastic thermodynamics \cite{seifert2012stochastic,derenyi1999efficiency,hondou2000unattainability,schmiedl2008,blickle2012realization}.
The analytic properties of the harmonic oscillator{---}in particular, the Gaussian form of the state{---}have motivated studies
of classical stochastic models of harmonic heat engines
\cite{raz2016geometric,blickle2012realization,dechant2016underdamped}. When comparing the two theories,
the results seem identical in many cases. Observing the Heisenberg equations of motion for the thermodynamical variables,
$\hbar$ does not appear.  Planck's constant in the commutators is cancelled by the inverse Planck constant in the equation of motion.
This raises the issue of what is  quantum in the quantum harmonic oscillator, or a~related issue---{\em what is quantum in quantum thermodynamics} \cite{vinjanampathy2015}?

In this review, we emphasized the power of the generalized Gibbs state in allowing a concise description
of an out-of-equilibrium situation of non-commuting operators. 
Using properties of Lie algebra of operators, we could obtain a dynamical description
of the state based on only  three variables: $\Op H$, $\Op L$, and $\Op C$. 
In the spirit of open quantum systems,
we could describe the cycle propagator as a~catenation of stroke propagators. 
All these propagators were cast
in the framework of the operator algebra, showing the power of Heisenberg representation.
The quantum variables were chosen to have direct thermodynamical relevance as energy and coherence.
In this review, we emphasized the connections between the algebraic approach and other popular
methods that have been employed to obtain insight on the harmonic engine. 

This formalism allows the cycles  to be classified according to the role of coherence. 
If the coherence vanishes at the
points where the strokes meet, frictionless cycles are obtained. 
Such cycles require special scheduling of $\omega(t)$
so that the coherence generated at the beginning of the stroke can be cashed upon at the end.
We reviewed the different approaches to obtain such scheduling 
and the minimum time that such moves can
be generated. This period is related to quantum speed limits \cite{mandelstam1945,margolus1998maximum,giovannetti2003,deffner2013,k310}, which are in turn related to the energy resources available to the system.
We chose the geometric mean $\tau_a \propto 1/\sqrt{ \omega_h \omega_c}$ to represent the minimum time allocation.
Faster scheduling requires unreasonable constraints on the stored energy in the oscillator during the stroke. 
We also assume that this extra energy required to achieve  the fast control is not dissipated
and can be accounted for as a catalyst.

For these frictionless solutions on the {\em adiabats},
the optimal time allocation for thermalisation is finite, leading to incomplete thermalisation. 
This allows  the minimum cycle time for frictionless cycles to be estimated.
Avoiding friction completely is an ideal that practically cannot be obtained. 
Using a~simple noise model, we show that some friction will always be present.

The model  demonstrates the fundamental tradeoff between efficiency and power. 
The frictionless solutions are a demonstration that quantum coherence which is related to friction can 
be cashed upon, using interference to cancel this friction. 
As a result, the maximum efficiency of the engine can be obtained in finite cycle time. 
Nevertheless, the Otto efficiency is smaller than the reversible Carnot efficiency
$\eta_o =1 -\frac{\omega_c}{\omega_h} \le \eta_c =1-\frac{T_c}{T_h}$, and operating at the Carnot efficiency will lead
to zero power. The entropy production can be associated with the heat transport, and for this case the entropy production
is linearly related to the power. Maximum power also implies maximum entropy production. 
This~finding is consistent with the study of Shiraishi et al. \cite{PhysRevLett.117.190601}. Any finite power
cycle requires out-of-equilibrium setups that lead to dissipation. In the sudden limit, there is no reversible choice.
Even at zero power the entropy production is positive. This could be the cost of maintaining~coherence.

Beyond a minimum time allocation on the {\em adiabats} $\tau_a$,  friction cannot be avoided. 
The transition point is the exceptional point of the non-hermitian degeneracy on the adiabatic propagator \cite{k282}. 
These~short time cycles are in the realm of the sudden cycles. 
The sudden cycles are an example of an~engine or refrigerator with no classical analogue. 
Power production requires coherence. 
A~sudden model without coherence operates as a dissipator generating entropy on both the hot and cold baths.
The sudden cycle is composed of non-commuting propagators with small action. 
Such cycles are universal and have a common continuous limit \cite{k299,k306}.
In the continuous limit, friction and heat leaks cannot be avoided \cite{correa2015internal,plastina2014}.

An obvious direction to look for quantum effects is to go to low temperatures where the  
unit of energy changes from $k_B T$  to $\hbar \omega$.
The adiabatic expansion is the bottleneck for cooling to extremely low temperatures.
The zero point energy plays an important role. We can approach the ground state on the hot side
by increasing the frequency, leading to the minimal initial energy $E_A=\frac{1}{2} \hbar \omega_h$.
This is a~large amount of energy compared to the cold side, which has to be eliminated adiabatically
or by using frictionless protocols. Any small error in these protocols will null the cooling.

The Otto quantum refrigerator is a good example for gaining insight into the limits of cooling
when operating at extremely low temperatures. 
Such refrigerators are an integral part of any quantum technology.
In the reciprocating Otto cycle, the cooling power is restricted either by the {\em adiabatic } expansion
or by vanishing of the heat transport when $T_c \rightarrow 0$ \cite{k275}. 
The {\em adiabatic } expansion time  is an~intrinsic property of the working medium. 
For optimal frictionless solutions, it scales as \linebreak $\tau_{hc} =O (T_c^{\frac{1}{2}})$, which gives a maximum rate of
entropy production $\sigma = O (T_c^{\frac{1}{2}})$, thus vanishing when $T_c \rightarrow 0$.
This is a demonstration of a dynamical version of the Nernst heat law \cite{k243,k275,k281}. 

The  quantum harmonic Otto cycle has been a template for many models
of quantum heat devices due to its analytic properties---for example, Otto cycles with  interacting particles \cite{wang2015efficiency,jaramillo2016quantum} or operating with many modes \cite{terccas2017quantum}.
The protocols developed for the harmonic case are generalised to eliminate friction in many-body dynamics. 

The quantum harmonic Otto cycle has been a source of inspiration for theory and experiment. 
The~model  incorporates generic features of irreversible operation which includes friction and  heat transport.
The system can bridge the conceptual gap between a single microscopic device to a~macroscopic heat~engine.

\vspace{6pt}

\acknowledgments{{We thank  Amikam Levy, Tova Feldmann, Raam Uzdin and Erik Torrontegui for sharing their ideas and insights.
We also thank Peter Salamon, Gonzales Muga, Robert Alicki for fruitful discussions.
We~acknowledge funding by the Israel Science Foundation, 
and COST Action MP1209 {\em Thermodynamics in the quantum regime}.}}

\bibliographystyle{mdpi}
\renewcommand\bibname{References}

\end{document}